\documentclass[10pt]{iopart}

\usepackage{graphicx} % standard LaTeX graphics for including eps-figure files
\usepackage{cite} %this package collapses a list of three or more consecutive reference numbers into a range!

\begin{document}

\newcommand{\dd}{{\rm{d}}} % diferencial tj. stojate d s argumentem
\newcommand{\rovno}{& = &} % rovnitko se zarovnanim pro pouztiti v eqnarray
\newcommand{\ssqrt}{{\textstyle\frac1{\sqrt{2}}}} % small inverse sqrt of 2
\newcommand{\df}{{\phi}} % diferencial tj. stojate f s argumentem

\newcommand{\boldu}{\mbox{\boldmath$u$}} % bold u
\newcommand{\bolde}{\mbox{\boldmath$e$}} % bold e
\newcommand{\boldk}{\mbox{\boldmath$k$}} % bold k
\newcommand{\boldl}{\mbox{\boldmath$l$}} % bold l
\newcommand{\boldm}{\mbox{\boldmath$m$}} % bold m
\newcommand{\bboldm}{\bar{\mbox{\boldmath$m$}}} % bar bold m
\newcommand{\boldZ}{\mbox{\boldmath$Z$}} % bold Z
\newcommand{\boldT}{\mbox{\boldmath$T$}} % bold T
\newcommand{\boldF}{\mbox{\boldmath$F$}} % bold F

\title[Explicit algebraic classification of Kundt geometries in any dimension]{Explicit algebraic classification of Kundt geometries in any dimension}

\author{J Podolsk\'y$^1$ and R \v{S}varc$^{1,2}$ }

\address{$^1$ Institute of Theoretical Physics, Faculty of Mathematics and Physics,
Charles University in Prague, V~Hole\v{s}ovi\v{c}k\'ach~2, 180~00 Praha 8, Czech Republic }
\address{$^2$ Department of Physics, Faculty of Science,
J.~E.~Purkinje University in {\'U}st{\'i} nad Labem, {\v C}esk{\'e} ml{\'a}de{\v z}e~8,
400~96 {\'U}st{\'i} nad Labem, Czech Republic}
\eads{\mailto{podolsky@mbox.troja.mff.cuni.cz} and \mailto{robert.svarc@mff.cuni.cz}}

\begin{abstract}
We present an algebraic classification, based on the null alignment properties of the Weyl tensor, of the general Kundt class of spacetimes in arbitrary dimension~$D$ for which the non-expanding, non-twisting, shear-free null direction~$\boldk$ is a (multiple) Weyl aligned null direction (WAND). No field equations are used, so that the results apply not only to Einstein's gravity and its direct extension to higher dimensions, but also to any metric theory of gravity which admits the Kundt spacetimes. By an explicit evaluation of the Weyl tensor in a natural null frame we demonstrate that all Kundt geometries are of type~I(b) or more special, and we derive simple necessary and sufficient conditions under which $\boldk$ becomes a double, triple or quadruple WAND. All possible algebraically special types, including the refinement to subtypes, are identified, namely II(a), II(b), II(c), II(d), III(a), III(b), N, O, II$_i$, III$_i$, D(a), D(b), D(c) and D(d). The corresponding conditions are surprisingly clear and expressed in an invariant geometric form. Some of them are always satisfied in four dimensions. To illustrate our classification scheme, we apply it to the most important subfamilies of the Kundt class, namely the pp-waves, the VSI spacetimes, and generalization of the Bertotti--Robinson, Nariai, and Pleba\'{n}ski--Hacyan direct-product spacetimes of any dimension.
\end{abstract}

\submitto{\CQG}

% 04.20.Jb Exact solutions
% 04.50.-h Higher-dimensional gravity and other theories of gravity
% 04.30.-w Gravitational waves: theory

\maketitle

\section{Introduction}
\label{intro}

Half a century ago, Wolfgang Kundt \cite{Kundt:1961,Kundt:1962} introduced and started to study one of the most important classes of exact spacetimes in Einstein's general relativity theory. These spacetimes are defined by a specific geometric property, namely that they admit a null geodesic congruence which is non-expanding, non-twisting, and shear-free, see chapter~31 of \cite{Stephani:2003} or chapter~18 of \cite{GriffithsPodolsky:2009}. Such a wide Kundt class admits various vacuum and pure radiation solutions, possibly with any value of the cosmological constant $\Lambda$, electromagnetic field, and other matter fields, including supersymmetry. These spacetimes may be of all algebraic types (namely of the Petrov type~N, III, D, II, I, or conformally flat). Interestingly, the Kundt geometries can also be explicitly extended to any number~$D$ of higher dimensions.

Among the famous subclasses of Kundt geometries (in four and higher dimensions) there are pp-waves which admit a covariantly constant null vector field \cite{Stephani:2003,GriffithsPodolsky:2009,Coley:2008,OrtaggioPravdaPravdova:2013,Bri25,ColMilPelPraPraZal03,ColeyMilsonPravdaPravdova:2004,ColMilPraPra04,ColFusHerPel06}. These include relativistic gyratons
\cite{Bon70,FroFur05,FroIsrZel05,FroZel05,FroZel06,KadlecovaZelnikovKrtousPodolsky:2009,KrtousPodolskyZelnikovKadlecova:2012} which represent the fields of localized spinning sources that propagate with the speed of light. The Kundt class also contains VSI and CSI spacetimes \cite{Pravdaetal:2002,ColMilPelPraPraZal03,ColeyMilsonPravdaPravdova:2004,Bri25,ColMilPraPra04,ColHerPel06,ColFusHerPel06,Coley:2008,ColHerPel09,OrtaggioPravdaPravdova:2013} for which all polynomial scalar invariants constructed from the Riemann tensor and its derivatives vanish and are constant, respectively.

In four dimensions, conformally flat pure radiation Kundt spacetimes provide an exceptional case for the invariant classification of exact solutions \cite{KoutrasMcIntosh:1996,EdgarLudwig:1997a,Skea:1997,GriffithsPodolsky:1998} and tests of the GHP and GIF formalisms \cite{EdgarVickers:1999,EdgarRamos:2007b,PodolskyPrikryl:2009}. All pure radiation type~D solutions are known \cite{GrooteBerghWylleman:2010}, and all electrovacuum Kundt solutions of type~D with an arbitrary cosmological constant $\Lambda$ were also found and studied
\cite{Carter:1968,Kinnersley:1969a,Plebanski:1975,PlebanskiDemianski:1976,Plebanski:1979,Garcia:1984,DebeverKamranMcLenaghan:1984,GriffithsPodolsky:2006b}. These contain a subfamily of direct-product spacetimes, namely the Bertotti--Robinson, (anti-)Nariai, and Pleba\'{n}ski--Hacyan spacetimes of type O and D, see chapter~7 of \cite{GriffithsPodolsky:2009}. Together with Minkowski and (anti-)de~Sitter spaces they form backgrounds on which non-expanding gravitational waves and gyratons of types~N and~II propagate \cite{OzsvathRobinsonRozga:1985,Siklos:1985,Podolsky:1998a,BicakPodolsky:1999a,BicakPodolsky:1999b,
GarciaAlvarez:1984, Khlebnikov:1986, Ortaggio:2002, GriffithsDochertyPodolsky:2004,PodolskyOrtaggio:2003,
KadlecovaZelnikovKrtousPodolsky:2009,KrtousPodolskyZelnikovKadlecova:2012}.
It is an interesting open problem to find and analyse possible extensions of all such spacetimes to higher dimensions.

The study of Kundt spacetimes is thus an active research area with a lot of applications, ranging from purely mathematical aspects to the investigation of various physical properties and models. In this paper we consider the fully general class of Kundt geometries in an arbitrary dimension ${D\ge4}$, without \emph{a priori} assuming any field equations. Specifically, we present their explicit and complete classification into the primary algebraic types and corresponding subtypes based on the WAND multiplicity of the optically privileged null vector field $\boldk$. We hope that the analysis may help us to understand the rich and interesting family of Kundt geometries and to elucidate mutual relations between its subclasses. The results apply to \emph{any} dimension, the algebraic classification within standard general relativity is simply obtained by setting ${D=4}$ and applying Einstein's field equations.

In section~\ref{sec_geom}, we start with fully general Kundt metric, introduce a suitable null frame, and project the Weyl tensor onto it. The corresponding Weyl scalars of all boost weights are employed for the algebraic classification.  In subsequent sections~\ref{claasifKundtI}--\ref{claasifKundtD} we derive the necessary and sufficient conditions for the specific algebraic types~II, III, N, O, D, including their subtypes. These are summarized in section~\ref{summarysub}.  The subclasses of pp-waves, VSI spacetimes, and generalized direct-product spacetimes are discussed in the final sections~\ref{ppwaves}--\ref{DObackgrounds}, respectively.
Explicit coordinate components of the Riemann, Ricci, and Weyl tensors for the generic Kundt geometry are given in \ref{appendixA}, and for the algebraically special Kundt spacetimes in \ref{appendixB}.

\section{The Weyl tensor of a general Kundt geometry}
\label{sec_geom}

In an arbitrary dimension $D$, the Kundt class is defined by admitting a non-expanding, twist-free, and shear-free null congruence. Such a geometric definition can be naturally expressed in terms of the optical scalars $\Theta$ (expansion), $A^2$ (twist) and $\sigma^2$ (shear) \cite{FroSto03,PraPraColMil04,Coley:2008,OrtaggioPravdaPravdova:2013} which, for affinely parameterized geodesic null congruence generated by a null vector field $\boldk$, are
\begin{equation}
\hspace{-15mm}
\Theta=\frac{1}{D-2}\,k^{a}_{\; ;a} \, , \quad A^2=-k_{[a;b]} k^{a;b} \, , \quad \sigma^2=k_{(a;b)} k^{a;b}-\frac{1}{D-2}\,(k^{a}_{\; ;a})^2 \, .\label{optical scalars}
\end{equation}
For the Kundt family ${\Theta=0}$, ${A=0}$ and ${\sigma=0}$, in which case there exist suitable coordinates such that any Kundt spacetime can be written as~\cite{Kundt:1961,Kundt:1962,Stephani:2003,PodOrt06,GriffithsPodolsky:2009,PodolskyZofka:2009,ColeyEtal:2009}
\begin{equation}
\hspace{-15mm}
\dd s^2 = g_{pq}(u,x)\, \dd x^p\dd x^q+2\,g_{up}(r,u,x)\, \dd u\,\dd x^p-2\,\dd u\,\dd r+g_{uu}(r,u,x)\,\dd u^2
\, . \label{obecny Kundtuv prostorocas}
\end{equation}
The coordinate $r$ is the affine parameter along the ``optically privileged'' null congruence (${\boldk=\partial_r}$), ${u=\ }$const. label null (wave)surfaces, and ${x\equiv(x^2, x^3, \ldots, x^{D-1})}$ are ${(D-2)}$ spatial coordinates in the transverse Riemannian space. Notice that the corresponding spatial part $g_{pq}$ of the metric must be independent of $r$, all other metric components $g_{up}$ and $g_{uu}$ can, in principle, be functions of all the coordinates $(r,u,x)$.

For such most general Kundt line element (\ref{obecny Kundtuv prostorocas}) the Christoffel symbols and the coordinate components of the Riemann, Ricci, and Weyl curvature tensors are presented in \ref{appendixA}.

As in \cite{ColeyMilsonPravdaPravdova:2004,Coley:2008,OrtaggioPravdaPravdova:2013}, algebraic classification of spacetimes here refers to determining specific properties of the Weyl tensor components (of different boost weights) with respect to a null frame ${\{\boldk, \boldl, \boldm_{i} \}}$ whose vectors satisfy the normalization conditions ${\boldk \cdot \boldl=-1}$, ${\boldm_i \cdot \boldm_j=\delta_{ij}}$, and ${\boldk \cdot \boldk=0=\boldl \cdot \boldl}$, ${\boldm_i \cdot \boldk=0=\boldm_i \cdot \boldl}$. In this work we will denote these Weyl tensor components as
\begin{eqnarray}
\Psi_{0^{ij}} \rovno C_{abcd}\; k^a\, m_i^b\, k^c\, m_j^d \, , \nonumber \\
\Psi_{1T^{i}} \rovno C_{abcd}\; k^a\, l^b\, k^c\, m_i^d \, ,
\hspace{10mm} \Psi_{1^{ijk}} = C_{abcd}\; k^a\, m_i^b\, m_j^c\, m_k^d \ ,\nonumber \\
\Psi_{2S} \rovno C_{abcd}\; k^a\, l^b\, l^c\, k^d \, ,
\hspace{11mm} \Psi_{2^{ijkl}}= C_{abcd}\; m_i^a\, m_j^b\, m_k^c\, m_l^d \, ,\nonumber \\
\Psi_{2T^{ij}}\rovno C_{abcd}\; k^a\, m_i^b\, l^c\, m_j^d \, ,
\hspace{10.2mm} \Psi_{2^{ij}} = C_{abcd}\; k^a\, l^b\, m_i^c\, m_j^d \, ,\nonumber \\
\Psi_{3T^{i}} \rovno C_{abcd}\; l^a\, k^b\, l^c\, m_i^d \, ,
\hspace{10.9mm} \Psi_{3^{ijk}} = C_{abcd}\; l^a\, m_i^b\, m_j^c\, m_k^d \, ,\nonumber\\
\Psi_{4^{ij}} \rovno C_{abcd}\; l^a\, m_i^b\, l^c\, m_j^d \, , \label{defPsiCoef}
\end{eqnarray}
where the indices ${\,i,j,k,l=2,\ldots,D-1\,}$ label the spatial Cartesian vectors $\boldm_{2}$, $\boldm_{3}$, $\ldots$, $\boldm_{D-1}$. The scalars (\ref{defPsiCoef}), listed by their boost weight, directly generalize the standard Newman--Penrose complex scalars $\Psi_A$ known from the ${D=4}$ case~\cite{KrtousPodolsky:2006,PodolskySvarc:2012}. All such scalars respect the standard symmetries of the Weyl tensor, for example
\begin{equation}
\Psi_{0^{ij}} = \Psi_{0^{(ij)}} \, ,\hspace{2.4mm} \Psi_{0^{k}}{}^{_k} = 0 \, ,\hspace{9.85mm}
\Psi_{4^{ij}} = \Psi_{4^{(ij)}} \, ,\hspace{2.4mm} \Psi_{4^{k}}{}^{_k} = 0 \, .\label{symmetries}
\end{equation}
There are relations between the scalars in the left and right columns of (\ref{defPsiCoef}), namely
\begin{eqnarray}
\Psi_{1T^i} \rovno \Psi_{1^{k}}{}^{_k}{}_{^i} \,,\nonumber \\
\Psi_{2S} \rovno\Psi_{2T^{k}}{}^{_k}\,,\quad
\Psi_{2T^{[ij]}}={\textstyle\frac{1}{2}}\Psi_{2^{ij}} \, ,\quad \Psi_{2T^{(ij)}}={\textstyle\frac{1}{2}}\Psi_{2^{ikj}}{}^{_k}\,,\label{constraints}\\
\Psi_{3T^i} \rovno \Psi_{3^{k}}{}^{_k}{}_{^i} \,.\nonumber
\end{eqnarray}
We should also emphasize that our notation, which in any dimension uses the symbols $\Psi_{A^{...}}$, is equivalent to the notations employed elsewehere, namely in \cite{ColeyMilsonPravdaPravdova:2004,Coley:2008}, \cite{PravdaPravdovaColeyMilson:2004,PraPraOrt07}, and \cite{DurPraPraReall10,OrtaggioPravdaPravdova:2013}. The identifications for the corresponding components are summarized in table~\ref{notationcomp}. The different signs arise from the opposite orientation of the null vector ${\boldl \rightarrow-\boldl}$ resulting in a different normalization condition ${\boldk \cdot \boldl=+1}$ (see \cite{PodolskySvarc:2012}).
\renewcommand{\arraystretch}{1.2}
\begin{table}[ht]
\begin{tabular}{|l||c|c|c|}
\hline
& ref.~\cite{ColeyMilsonPravdaPravdova:2004,Coley:2008} & ref.~\cite{PravdaPravdovaColeyMilson:2004,PraPraOrt07} & ref.~\cite{DurPraPraReall10,OrtaggioPravdaPravdova:2013} \\
\hline\hline
$\Psi_{0^{ij}} $ & $ C_{0i0j}$ & & $\Omega_{ij}$ \\
\hline
$\Psi_{1T^{j}} $ & $-C_{010j}$ & & $-\Psi_j$ \\
\hline
$\Psi_{1^{ijk}}$ & $C_{0ijk}$ & & $\Psi_{ijk}$ \\
\hline
$\Psi_{2S}$ & $-C_{0101}$ & $-\Phi$ & $-\Phi$ \\
\hline
$\Psi_{2T^{ij}}$ & $-C_{0i1j}$ & $-\Phi_{ij}$ & $-\Phi_{ij}$ \\
\hline
$\Psi_{2T^{(ij)}}$& & & $-\Phi^{\rm S}_{ij}$ \\
\hline
$\Psi_{2T^{[ij]}}$& & & $-\Phi^{\rm A}_{ij}$ \\
\hline
$\Psi_{2^{ij}} $ & $-C_{01ij}$ & & $-2\,\Phi^{\rm A}_{ij}$ \\
\hline
$\Psi_{2^{ijkl}}$ & $ C_{ijkl}$ & & $\Phi_{ijkl}$ \\
\hline
$\Psi_{3T^{j}}$ & $ C_{101j}$ & $\Psi_j$ & $\Psi'_j$ \\
\hline
$\Psi_{3^{ijk}}$ & $-C_{1ijk}$ & & $-\Psi'_{ijk}$ \\
\hline
$\Psi_{4^{ij}}$ & $ C_{1i1j}$ & $2\,\Psi_{ij}$& $\Omega'_{ij}$ \\
\hline
\end{tabular}
%\vspace{3mm}
\caption{\label{notationcomp}Different equivalent notations used in the literature for the Weyl scalars, in particular the GHP formalism~\cite{DurPraPraReall10,OrtaggioPravdaPravdova:2013}.}
\end{table}

To evaluate all the scalars (\ref{defPsiCoef}) for the general Kundt spacetime~(\ref{obecny Kundtuv prostorocas}), we project the coordinate components of the Weyl tensor (\ref{C_abcd}) onto the natural null frame
\footnote{In this paper, ${i,j,k,l}$ are frame labels, whereas the indices ${p,q,m,n}$ denote the spatial coordinate components. For example, $m_{i}^{p}$ stands for the $p^{\hbox{\tiny{th}}}$  spatial coordinate component of the vector $\boldm_i$.}
\begin{eqnarray}
\boldk \rovno \, \mathbf{\partial}_r \ , \nonumber \\
\boldl\rovno \, {\textstyle\frac{1}{2}}g_{uu}\,\mathbf{\partial}_r+\mathbf{\partial}_u\, , \label{Kundt simplest frame}\\
\boldm_i\rovno \, m_{i}^{p}\,(\,g_{up}\,{\partial}_r+\mathbf{\partial}_p) \, , \nonumber
\end{eqnarray}
where the coefficients ${\,m_i^p\,}$ satisfy ${{\,g_{pq}\,m_i^p \,m_j^q}=\delta_{ij}}$ to fulfil the normalization conditions ${\boldm_i \cdot \boldm_j=\delta_{ij}}$ and ${\boldk \cdot \boldl=-1}$. The vector $\boldk$ is oriented along the optically privileged null congruence ${\partial_r}$ defining the Kundt family. Direct calculation yields
\begin{eqnarray}
&&\hspace{-25mm}
\Psi_{0^{ij}} \hspace{1.7mm}= 0 \,, \label{Psi 0ij - rozepsany frame} \\
&&\hspace{-25mm}
\Psi_{1T^j} \hspace{0.9mm}= m_{j}^{p}\, C_{rpru} \, ,\label{Psi 1Tj - rozepsany frame} \\
&&\hspace{-25mm}
\Psi_{2S} \hspace{2.5mm}= -C_{ruru} \, ,\label{Psi 2S - rozepsany frame} \\
&&\hspace{-25mm}
\Psi_{2T^{ij}} = m_{i}^{p}m_{j}^{q}\,(-C_{rpru}g_{uq}+C_{rpuq})\,, \label{Psi 2Tij - rozepsany frame} \\
&&\hspace{-25mm}
\Psi_{3T^j} \hspace{1.12mm} = m_{j}^{p}\,\big(\!-\textstyle{\frac{1}{2}}C_{rpru}g_{uu}+C_{ruru}g_{up}-C_{ruup}\big)\,, \label{Psi 3Tj - rozepsany frame} \\
&&\hspace{-25mm}
\Psi_{4^{ij}} \hspace{2.4mm} = m_{i}^{p}m_{j}^{q}\,\big[\,C_{ruru}g_{up}g_{uq}-\textstyle{\frac{1}{2}}g_{uu}(C_{rpru}g_{uq}+C_{rqru}g_{up})
\nonumber\\
&&\hspace{-2mm}
+\textstyle{\frac{1}{2}}g_{uu}(C_{rpuq}+C_{rqup})-(C_{ruup}g_{uq}+C_{ruuq}g_{up})+C_{upuq}\,\big], \label{Psi 4ij - rozepsany frame}\\
&&\hspace{-25mm}
\Psi_{1^{ijk}} \hspace{1.05mm} = m_{i}^{p}m_{j}^{m}m_{k}^{q}\,C_{rpmq} \, , \label{Psi 1ijk - rozepsany frame}\\
&&\hspace{-25mm}
\Psi_{2^{ij}} \hspace{2.3mm} = m_{i}^{p}m_{j}^{q}\,(C_{rqru}g_{up}-C_{rpru}g_{uq}+C_{rupq})\,, \label{Psi 2ij - rozepsany frame}\\
&&\hspace{-25mm}
\Psi_{2^{ijkl}} = m_{i}^{m}m_{j}^{p}m_{k}^{n}m_{l}^{q}\,(C_{rpnq}g_{um}-C_{rmnq}g_{up}+C_{rqmp}g_{un}-C_{rnmp}g_{uq}+C_{mpnq}), \label{Psi 2ijkl - rozepsany frame}\\
&&\hspace{-25mm}
\Psi_{3^{ijk}} \hspace{0.95mm}= m_{i}^{p}m_{j}^{m}m_{k}^{q}\,\big(\textstyle{\frac{1}{2}}C_{rpmq}g_{uu}+C_{rmru}g_{up}g_{uq}-C_{rqru}g_{up}g_{um} \nonumber \\
&& \hspace{11mm}+C_{rqup}g_{um}-C_{rmup}g_{uq}-C_{rumq}g_{up}+C_{upmq}\big)\,.\label{Psi 3ijk - rozepsany frame}
\end{eqnarray}
To determine the specific algebraic type of any of the Kundt spacetimes we have to investigate the values and mutual relations of these scalars.

It is well known that for the general Kundt geometry (\ref{obecny Kundtuv prostorocas}) the privileged null vector ${\boldk=\partial_r}$ is a~WAND, so that the spacetime is of \emph{algebraic type}~I or more special \cite{OrtaggioPravdaPravdova:2007,PodolskyZofka:2009,OrtaggioPravdaPravdova:2013}. Indeed, we immediately see from (\ref{Psi 0ij - rozepsany frame}) that ${\Psi_{0^{ij}}=0}$, i.e., the Weyl tensor components of the highest boost weight ${+2}$ are completely missing.

Using the frame (\ref{Kundt simplest frame}) the boost weight~${+1}$ scalars
${\Psi_{1T^j}}$ and ${\Psi_{1^{ijk}}}$ for all frame labels ${i,j,k=2,\ldots,D-1}$ are, in view of (\ref{Psi 1Tj - rozepsany frame}), (\ref{Psi 1ijk - rozepsany frame}) and (\ref{C_abcd}), (\ref{Riemann rpru - Kundt}), (\ref{Ricci rk - Kundt}), explicitly
\begin{eqnarray}
\Psi_{1T^j} \rovno -\frac{1}{2}\,\frac{D-3}{D-2}\,m_{j}^{p} \,g_{up,rr} \,, \label{Psi1a}\\
\Psi_{1^{ijk}} \rovno \frac{1}{D-3}\,(\delta_{ij}\,\Psi_{1T^k}-\delta_{ik}\,\Psi_{1T^j}) \, .\label{Psi1b}
\end{eqnarray}
Clearly, ${\Psi_{1T^k}= \Psi_{1^{i}}{}^{_i}{}_{^k}}$ which confirms (\ref{constraints}). Moreover, we observe that there are no Kundt spacetimes of genuine subtype I(a), see the definition in section~2.3 of~\cite{OrtaggioPravdaPravdova:2013}. We thus proved that \emph{all Kundt geometries are actually of algebraic type}~I(b)$\equiv$I, or more special.

\section{Algebraically special (type II) Kundt spacetimes}
\label{claasifKundtI}

Kundt spacetimes are \emph{algebraically special}, that is of type~II (or more special), if the corresponding Weyl scalars $\Psi_{A^{...}}$ of the  boost weight ${+1}$ (or lower) also vanish.

In this paper we restrict ourselves to the geometrically privileged large subclass\footnote{There may exist ``non-aligned'' Kundt geometries for which the WAND $\boldk$ (oriented along the geodesic, non-expanding, non-twisting and shear-free congruence) is \emph{not} multiple while they admit \emph{another} non-Kundt WAND that \emph{is} multiple, cf.\cite{ColeyEtal:2009}.} of Kundt spacetimes which are of type~II or more special \emph{with respect to the WAND}~$\boldk$.

From (\ref{Psi1a}), (\ref{Psi1b}) it follows that the simple condition ${\,\Psi_{1T^j}=0\,}$, that is explicitly ${m_{j}^{p} \,g_{up,rr}=0}$, is the necessary and sufficient condition for a Kundt spacetime of any dimension to be of type~II$\equiv$I(a), i.e., algebraically special with double WAND~${\boldk=\partial_r}$. Since the $(D-2)$ spatial vectors $\boldm_j$ of (\ref{Kundt simplest frame}) forming the local Cartesian frame in the transverse Riemannian space are linearly independent, ${g_{up}}$ must be of the form
\begin{equation}
g_{up}=e_p(u,x)+ f_p(u,x) \,r\, , \label{guiprotypII}
\end{equation}
where for any ${p=2,\ldots,D-1\,}$ the functions ${e_p, f_p}$ are independent of the coordinate~$r$, confirming the results of \cite{OrtaggioPravdaPravdova:2013,ColeyEtal:2009}. \emph{The most general Kundt spacetime of algebraic type}~II \emph{with respect to double WAND}~$\boldk$ in any dimension $D$ can be thus written as
\begin{equation}
\hspace{-15mm}
\dd s^2 = g_{pq} \,\dd x^p\dd x^q+2\,(e_p+ f_p \,r)\,\dd u\,\dd x^p -2\,\dd u\,\dd r+g_{uu}(r,u,x)\,\dd u^2
\, , \label{obecny Kundt II}
\end{equation}
where $g_{pq}$, $e_p$, $f_p$ are functions of $u$ and $x$ only. Any Kundt spacetime of Weyl type~II with respect to ${\boldk}$ \emph{necessarily} implies ${g_{up,rr}=0}$ and thus ${R_{rp}=0}$, see (\ref{Ricci rk - Kundt}), which is equivalent to ${R_{ab}k^b\propto k_a}$. Consequently, the corresponding Ricci tensor must be of type~II, or more special, without any additional geometrical or physical condition (see Proposition 7.1 of \cite{OrtaggioPravdaPravdova:2013}).

For the algebraically special Kundt geometries with multiple WAND $\boldk$, we can evaluate all the remaining Weyl scalars of boost weights ${0,-1,-2}$. First, after substituting (\ref{guiprotypII}) into (\ref{Riemann rprq - Kundt})--(\ref{C_abcd}), a straightforward but very lengthy calculation yields the explicit forms of the Riemann, Ricci, and Weyl tensors for an \emph{arbitrary such algebraically special Kundt spacetime} of dimension ${D\ge4}$. The complete results are presented in \hbox{\ref{appendixB}}. Next, using the expressions (\ref{Psi 2S - rozepsany frame})--(\ref{Psi 3ijk - rozepsany frame})
and the Weyl tensor coordinate components (\ref{Weyl_rprqII})--(\ref{Weyl_upuqII}), after the surprising and non-trivial cancelation of various terms we obtain the following explicit expressions:
\begin{eqnarray}
&&\hspace{-20mm}\Psi_{2S} = \frac{D-3}{D-1}\left[\frac{1}{2}\,g_{uu,rr}-\frac{1}{4}f^p f_p+\frac{1}{D-2}\left(\frac{\,^{S}\!R}{D-3}+f\right)\right], \label{Psi2s}\\
&&\hspace{-20mm}
\Psi_{2T^{ij}} = \frac{g_{pq}\,m_{i}^{p}m_{j}^{q}}{(D-1)(D-2)}
\left[\frac{1}{2}(D-3)\,g_{uu,rr}-\frac{1}{4}(D-3)f^m f_m-\,^{S}\!R -\frac{1}{2}(D-5)f\,\right] \nonumber \\
&& \hspace{-6.4mm}
+\frac{1}{D-2}\,m_{i}^{p}m_{j}^{q} \left[\,^{S}\!R_{pq}+\frac{1}{2}(D-4)\,f_{pq}
+\frac{1}{2}(D-2)\,F_{pq}\right],\label{Psi2Tij}\\
&&\hspace{-20mm}
\Psi_{2^{ij}}= m_{i}^{p}m_{j}^{q}\,F_{pq}\,,\label{Psi2ij}\\
&&\hspace{-20mm}
\Psi_{2^{ijkl}}= m_{i}^{m}m_{j}^{p}m_{k}^{n}m_{l}^{q}
\bigg[
\,^{S}\!R_{mpnq}-\frac{1}{D-2}\Big(g_{mn}(\,^{S}\!R_{pq}-f_{pq})-g_{mq}(\,^{S}\!R_{pn}-f_{pn}) \nonumber \\
&& \hspace{33mm}+g_{pq}(\,^{S}\!R_{mn}-f_{mn})-g_{pn}(\,^{S}\!R_{mq}-f_{mq})\Big)\nonumber\\
&&\hspace{-6.4mm}+\frac{1}{(D-1)(D-2)}\Big(\,g_{uu,rr}+\,^{S}\!R-2f-{\textstyle\frac{1}{2}}f^s f_s \Big)(g_{mn}g_{pq}-g_{mq}g_{pn})\bigg],\label{Psi2ijkl}\\
&&\hspace{-20mm} \Psi_{3T^j} = -m_{j}^{p}\,\frac{D-3}{D-2}\,\bigg[\,\frac{1}{2} \Big(rf_p\,g_{uu,rr}
+g_{uu,rp}-f_{p,u}\Big)+e_p\Big(\frac{1}{2}\,g_{uu,rr}-\frac{1}{4}f^qf_q\Big) \nonumber \\
&&\hspace{1.4mm}+\frac{1}{4}f^qe_qf_p-\frac{1}{2}f^qE_{qp}-\frac{1}{D-3}\,X_p
-r\,\Big(\frac{1}{2}f^q F_{qp}+\frac{1}{D-3}Y_p\Big)\bigg],\label{Psi 3TjD}\\
&&\hspace{-20mm} \Psi_{3^{ijk}} = \frac{1}{D-3}\,(\delta_{ij}\,\Psi_{3T^k}-\delta_{ik}\,\Psi_{3T^j})+\tilde\Psi_{3^{ijk}} \, , \label{Psi3D}\\
&&\hspace{-20mm} \tilde\Psi_{3^{ijk}} = m_{i}^{p}m_{j}^{m}m_{k}^{q}\,
\bigg[\,\Big(X_{pmq}-\frac{1}{D-3}(\, g_{pm}\,X_q-g_{pq}\,X_m)\Big)\nonumber\\
&&\hspace{6mm}   +r\,\Big(Y_{pmq}-\frac{1}{D-3}(\, g_{pm}\,Y_q-g_{pq}\,Y_m)\Big)\,\bigg] \, , \label{Psi3ijkD}\\
&&\hspace{-20mm} \Psi_{4^{ij}} =  m_{i}^{p}m_{j}^{q}\, \bigg[\,W_{pq}-\frac{1}{D-2}\, g_{pq}\,W\,\bigg]  \, , \label{Psi4D}
\end{eqnarray}
\vspace{1.3mm}
where ${{\,g_{pq}\,m_i^p \,m_j^q}=\delta_{ij}\,}$ and
\begin{eqnarray}
&&\hspace{-16.3mm}
X_{pmq} \equiv
e_{[q||m]||p}}+F_{qm}\,e_p + {\textstyle\frac{1}{2}\!\left(F_{pm}\,e_q-F_{pq}\,e_m\right)  \nonumber\\
&&\hspace{-4mm}+{\textstyle \frac{1}{2}\!\left(e_{pm}f_q-e_{pq}f_m\right)-\frac{1}{2}\!\left(f_{pm}\,e_q-f_{pq}\,e_m\right)}+g_{p[m,u||q]}\, , \label{Xipj*} \\
&&\hspace{-16.3mm}
Y_{pmq} \equiv
f_{[q||m]||p}+ F_{qm}\,f_p + {\textstyle \frac{1}{2}}\!\left(F_{pm}\,f_q-F_{pq}\,f_m\right) , \label{YpmgD}\\
&&\hspace{-15mm}
W_{pq} \equiv {\textstyle
-\frac{1}{2}(g_{uu})_{||p||q}+\frac{1}{2}g_{uu}f_{(p||q)}+\frac{1}{4}(g_{uu,p}f_q+g_{uu,q}f_p)} \nonumber\\
&&\hspace{-4mm} {\textstyle
-\frac{1}{2}g_{uu,r}(r f_{(p||q)}+e_{pq})-\frac{1}{2}g_{uu,rr}(r^2 f_p f_q+r(f_p e_q+f_qe_p)+e_pe_q)}
\nonumber\\
&&\hspace{-4mm} {\textstyle
+\frac{1}{2}[(f_{p,u}-g_{uu,rp})(r f_q+e_q)+(f_{q,u}-g_{uu,rq})(r f_p+e_p)]}
\nonumber\\
&&\hspace{-4mm} {\textstyle
+ r^2 g^{mn}F_{mp}F_{nq} + r\Big[f_{(p,u||q)}+g^{mn}(E_{mp}F_{nq}+E_{mq}F_{np}) } \nonumber\\
&&\hspace{7mm} {\textstyle
+ \frac{1}{2}[f^m(F_{mp}\,e_q+F_{mq}\,e_p)-e^m(F_{mp}\,f_q+F_{mq}\,f_p)]\Big] } \nonumber\\
&&\hspace{-4mm} {\textstyle
+ (e_{(p,u||q)}-\frac{1}{2}g_{pq,uu})+g^{mn}E_{mp}E_{nq} }\nonumber\\
&&\hspace{7mm} {\textstyle
+ \frac{1}{2}[f^m(E_{mp}\,e_q+E_{mq}\,e_p)-e^m(E_{mp}\,f_q+E_{mq}\,f_p)]} \nonumber\\
&&\hspace{7mm} {\textstyle
+\frac{1}{4}(e^me_mf_pf_q+f^mf_me_pe_q)-\frac{1}{4}f^me_m(f_pe_q+f_qe_p)  }
\, . \label{WpqD}
\label{Yipj*}
\end{eqnarray}
The corresponding contractions ${\,X_q \equiv g^{pm}X_{pmq}\,}$, ${\,Y_q \equiv g^{pm}\,Y_{pmq}\,}$, ${\,W\equiv g^{pq}\,W_{pq}\,}$ are
\begin{eqnarray}
X_q &=& g^{pm}e_{[q||p]||m}-{\textstyle \frac{3}{2}} F_{pq}\, e^p   \nonumber \\
&&\hspace{0mm}
+{\textstyle \frac{1}{2}\!\left(f_{pq}\,e^p-e_{pq}f^p\right)+\frac{1}{2}\!\left(g^{pm}e_{pm}f_q-f\,e_q\right)+g^{pm}g_{p[m,u||q]}}\, , \label{Xp*}\\
Y_q   &=& g^{pm}f_{[q||p]||m}-{\textstyle \frac{3}{2}} F_{pq}\, f^p\, , \label{Yp*}\\
W     &=&
-{\textstyle\frac{1}{2}\triangle g_{uu}+\frac{1}{2}g_{uu}\,{f^p}_{||p}+\frac{1}{2}g_{uu,p}f^p} \nonumber\\
&&
-{\textstyle\frac{1}{2}g_{uu,r}\,g^{pq}(r f_{(p||q)}+e_{pq})-\frac{1}{2}g_{uu,rr}(r^2 f^p f_p+2\,r f^p e_p+e^pe_p)}
\nonumber\\
&&
+(f_{p,u}-g_{uu,rp})(r f^p+e^p)+ r^2 g^{pq}g^{mn}F_{pm}F_{qn}
\nonumber\\
&&
+ r\left[\,g^{pq}f_{(p,u||q)}+2\,g^{pq}g^{mn}E_{pm}F_{qn}+ 2\,f^pe^qF_{pq}\,\right]  \nonumber\\
&&
+ g^{pq}(e_{(p,u||q)}-{\textstyle\frac{1}{2}}g_{pq,uu})+g^{pq}g^{mn}E_{pm}E_{qn}+ 2\,f^pe^qe_{[p||q]} \nonumber\\
&&
+{\textstyle\frac{1}{2}e^pe_pf^qf_q-\frac{1}{2}f^pe_pf^qe_q } \, . \label{W}
\end{eqnarray}
We have introduced the convenient geometric quantities
\begin{eqnarray}
f^p &\equiv& g^{pq}f_q\,,\label{defkontra}\\
f_{p||q} &\equiv& f_{p,q}-\Gamma^m_{pq} f_m \,, \label{defderiv}\\
{f^p}_{||p} &\equiv& g^{pq}f_{p||q}\,, \label{defdiv}\\
f_{pq} &\equiv& f_{(p||q)} + {\textstyle\frac{1}{2}} f_p f_q\,, \label{deffij}\\
f &\equiv& g^{pq}f_{pq}={f^p}_{||p}+ {\textstyle\frac{1}{2}} f^p f_p\,, \label{deff}\\
F_{pq} &\equiv&  f_{[p||q]} = f_{[p,q]} \,, \label{deffija}\\
f_{[m||q]||p} &\equiv& f_{[m,q],p}-\Gamma^{n}_{pm}f_{[n,q]}-\Gamma^{n}_{pq}f_{[m,n]}\,, \label{defidef1t}\\
f_{p,u||q} &\equiv& (f_{p,u})_{||q}= f_{p,uq}-f_{n,u}\,\Gamma^{n}_{pq}\,, \label{defideeu11t}\\
f_{(p,u||q)} &\equiv& f_{(p,q),u}-f_{n,u}\,\Gamma^{n}_{pq}\,, \label{defideeu1t}\\
e^p &\equiv& g^{pq}e_q\,,\label{defkontrae}\\
e_{p||q} &\equiv& e_{p,q}-\Gamma^m_{pq} e_m \,, \label{defderivapet}\\
e_{pq} &\equiv& e_{(p||q)} - {\textstyle\frac{1}{2}} g_{pq,u}\,, \label{defeijapt}\\
E_{pq} &\equiv& e_{[p||q]}\,+{\textstyle \frac{1}{2}}g_{pq,u}\,, \label{defEijapt}\\
e_{[m||q]||p} &\equiv& e_{[m,q],p}-\Gamma^{n}_{pm}e_{[n,q]}-\Gamma^{n}_{pq}e_{[m,n]}\,, \label{defidee1t}\\
e_{p,u||q} &\equiv& (e_{p,u})_{||q}= e_{p,uq}-e_{n,u}\,\Gamma^{n}_{pq}\,, \label{defideeu22t}\\
e_{(p,u||q)} &\equiv& e_{(p,q),u}-e_{n,u}\,\Gamma^{n}_{pq}\,, \label{defideeu2t}
\end{eqnarray}
and
\begin{eqnarray}
g_{p[m,u||q]} &\equiv& {\textstyle \frac{1}{2}}[\,(g_{pm,u})_{||q}-(g_{pq,u})_{||m}\,] \label{defideeu28t}\\
&&\hspace{-4.6mm}
 =  g_{p[m,q],u}+{\textstyle \frac{1}{2}}(\,\Gamma^{n}_{pm}\,g_{nq,u}-\Gamma^{n}_{pq}\,g_{nm,u}) \,, \nonumber\\
(g_{uu})_{||p||q} &\equiv& g_{uu,pq}-g_{uu,n}\,\Gamma^{n}_{pq}\,, \label{defideeu3t}\\
\triangle g_{uu} &\equiv& g^{pq}(g_{uu})_{||p||q}\,, \label{Laplaceguut}
\end{eqnarray}
in which the symbol $||$ indicates the covariant derivative with respect to the spatial metric $g_{pq}$ in the transverse ${(D-2)}$-dimensional Riemannian space, whose metric is $g_{pq}$ and the corresponding Riemann and Ricci curvature tensors are ${\,^{S}\!R_{mpnq}}$ and ${\,^{S}\!R_{pq}}$, respectively.

We should emphasize that all the quantities (\ref{defkontra})--(\ref{defideeu28t}) are \emph{independent} of the coordinate~$r$. It is also important to observe that ${\,W_{pq} = W_{qp}\,}$, while ${\,X_{pmq} = -X_{pqm}\,}$, ${\,Y_{pmq} = -Y_{pqm}}$. Consequently, ${X_q \equiv g^{pm}X_{pmq}}$ and ${Y_q \equiv g^{pm}Y_{pmq}}$, given in (\ref{Xp*}), (\ref{Yp*}),  are the only non-trivial contractions of $X_{pmq}$ and $Y_{pmq}$, respectively.

By taking the symmetric and antisymmetric parts of $\Psi_{2T^{ij}}$ given by expression (\ref{Psi2Tij}) we immediately obtain
\begin{eqnarray}
&&\hspace{-23mm}
\Psi_{2T^{(ij)}} = \frac{g_{pq}\,m_{i}^{p}m_{j}^{q}}{(D-1)(D-2)}
\left[\frac{1}{2}(D-3)\,g_{uu,rr}-\frac{1}{4}(D-3)f^m f_m-\,^{S}\!R -\frac{1}{2}(D-5)f\,\right] \nonumber \\
&& \hspace{-7.5mm}
+\frac{1}{D-2}\,m_{i}^{p}m_{j}^{q} \left[\,^{S}\!R_{pq}+\frac{1}{2}(D-4)\,f_{pq}\right],\label{Psi2TijSYM}\\
&&\hspace{-23mm}
\Psi_{2T^{[ij]}} = \frac{1}{2}\, m_{i}^{p}m_{j}^{q}\,F_{pq}\,,\label{Psi2TijANTI}
\end{eqnarray}
which explicitly confirm the relations ${\Psi_{2S}=\Psi_{2T^{k}}{}^{_k}\equiv\Psi_{2T^{(ij)}}\,\delta^{ij}}$ and ${\Psi_{2T^{[ij]}}={\textstyle\frac{1}{2}}\Psi_{2^{ij}}}$, see (\ref{constraints}). For the corresponding irreducible components defined as
\begin{eqnarray}
&&\hspace{-18mm}
\tilde\Psi_{2T^{(ij)}}\equiv\Psi_{2T^{(ij)}}-\frac{1}{D-2}\,\delta_{ij}\Psi_{2S}\,, \label{deftildePsi2TijSYM}\\
&&\hspace{-18mm}
\tilde\Psi_{2^{ijkl}}\equiv\Psi_{2^{ijkl}}-\frac{2}{D-4}\,(\delta_{ik}\Psi_{2T^{(lj)}}-\delta_{il}\Psi_{2T^{(kj)}}
-\delta_{jk}\Psi_{2T^{(li)}}+\delta_{jl}\Psi_{2T^{(ki)}})   \nonumber\\
&&\hspace{4.5mm}
+\frac{2}{(D-3)(D-4)}\,\Psi_{2S}\,(\delta_{ik}\delta_{lj}-\delta_{il}\delta_{kj})\,,\label{deftildePsi2ijkl}
\end{eqnarray}
we then get using (\ref{Psi2s}) and (\ref{Psi2ijkl}) that
\begin{eqnarray}
&&\hspace{-18mm}
\tilde\Psi_{2T^{(ij)}} = \frac{m_{i}^{p}m_{j}^{q}}{D-2}
\left[\left(\!\,^{S}\!R_{pq}-\frac{g_{pq}}{D-2}\,^{S}\!R\right)
+\frac{1}{2}(D-4)\left(f_{pq}-\frac{g_{pq}}{D-2}f\right)\right], \label{tildePsi2TijSYM}\\
&&\hspace{-18mm}
\tilde\Psi_{2^{ijkl}} = m_{i}^{m}m_{j}^{p}m_{k}^{n}m_{l}^{q}\,\,^{S}\!C_{mpnq} \,.\label{tildePsi2ijkl}
\end{eqnarray}

In general, all these Weyl scalars of boost weight 0 are non-trivial. However, if some of them vanish, we obtain an \emph{explicit refinement} of the algebraic classification of type~II Kundt spacetimes with double WAND~${\boldk=\partial_r}$ into the corresponding \emph{subtypes} defined in~\cite{OrtaggioPravdaPravdova:2013} (see also \cite{ColeyMilsonPravdaPravdova:2004,Ortaggio:2009,ColeyHervik:2010}). To be specific:

\bigskip

\begin{itemize}

\item
The Kundt spacetime (\ref{obecny Kundt II}) is of \emph{algebraic type}~II(a) ${\Leftrightarrow\Psi_{2S} =0 \Leftrightarrow}$ the metric function $g_{uu}$ is \emph{at most quadratic} in the coordinate~$r$,
\begin{equation}
g_{uu}=a(u,x)\,r^2+ b(u,x)\,r+c(u,x)\, , \label{guuprospinttypII}
\end{equation}
and $a$ is uniquely given as
\begin{eqnarray}
a&=& \frac{1}{4}f^p f_p-\frac{1}{D-2}\left(\frac{\,^{S}\!R}{D-3}+f\right),
\label{typIIa}
\end{eqnarray}
where ${^{S}\!R}$ is the Ricci scalar of the spatial metric $g_{pq}$, and $f$ is defined by~(\ref{deff}).
Notice that these spacetimes form a subclass of the degenerate Kundt spacetimes studied in \cite{ColeyEtal:2009}.

\item
The Kundt spacetime (\ref{obecny Kundt II}) is of \emph{algebraic type}~II(b) ${\Leftrightarrow\tilde\Psi_{2T^{(ij)}} =0 \Leftrightarrow}$
\begin{equation}
\,^{S}\!R_{pq}-\frac{g_{pq}}{D-2}\,^{S}\!R
=-\frac{1}{2}(D-4)\left(f_{pq}-\frac{g_{pq}}{D-2}f\right). \label{typIIb}
\end{equation}
This is identically satisfied when ${D=4}$ since for any transverse 2-dimensional Riemannian space there is ${\,^{S}\!R_{pq}=\frac{1}{2}g_{pq}\,^{S}\!R}$.

\item
The Kundt spacetime (\ref{obecny Kundt II}) is of \emph{algebraic type}~II(c) ${\Leftrightarrow\tilde\Psi_{2^{ijkl}} =0 \Leftrightarrow}$
\begin{equation}
\,^{S}\!C_{mpnq}=0 \, . \label{typIIc}
\end{equation}
This is always satisfied when ${D=4}$ and ${D=5}$ since the Weyl tensor vanishes identically in dimensions 2 and 3, see also \cite{ColeyHervik:2010,ColeyEtal:2012}.

\item
The Kundt spacetime (\ref{obecny Kundt II}) is of \emph{algebraic type}~II(d) ${\Leftrightarrow\Psi_{2T^{[ij]}} =0 \Leftrightarrow}$
\begin{equation}
F_{pq}=0\, , \label{typIId}
\end{equation}
i.e., ${f_{p,q}-f_{q,p}=0}$ for all ${p,q}$, see (\ref{deffija}).
This condition can be conveniently rewritten in a geometric form. Introducing the 1-form ${\df\equiv f_p\,\dd x^p}$ in the transverse ${(D-2)}$-dimensional Riemannian space, (\ref{typIId}) is equivalent to the condition that $\df$ is \emph{closed}, ${\dd \df=0}$. On any \emph{contractible domain}, every closed form is exact by the Poincar\'{e} lemma, so that there exists a \emph{potential function} ${\cal F}$ such that ${\df=\dd{\cal F}}$,
\begin{equation}
f_p\equiv{\cal F}_{,p}\, . \label{fpotential}
\end{equation}
In a general case, such ${\cal F}$ exists only \emph{locally}.

\end{itemize}

\section{Type III Kundt spacetimes}
\label{claasifKundtIII}

The Kundt spacetime (\ref{obecny Kundt II}) is of \emph{algebraic type}~III with respect to the triple WAND ${\boldk=\partial_r}$ if all these four independent conditions (\ref{guuprospinttypII})--(\ref{typIId}) are \emph{satisfied simultaneously}, that is III$\equiv$II(abcd). Recall that in four spacetime dimensions, the conditions (\ref{typIIb}) and (\ref{typIIc}) are identically valid. For type~III Kundt spacetimes the general expressions (\ref{Psi 3TjD}), (\ref{Psi3D}), (\ref{Psi3ijkD}) for the corresponding Weyl scalars of the boost weight ${-1}$ thus reduce to expressions
\begin{eqnarray}
&&\hspace{-8mm} \Psi_{3T^j} = -m_{j}^{p}\,\frac{D-3}{D-2}\,\bigg[(a_{,p}+f_{p}\,a)\,r +\frac{1}{2}(b_{,p}-f_{p,u})-\frac{1}{2}\,T_p\bigg]
\, , \label{Psi 3Tj}\\
&&\hspace{-8mm} \Psi_{3^{ijk}} = \frac{1}{D-3}\,(\delta_{ij}\,\Psi_{3T^k}-\delta_{ik}\,\Psi_{3T^j})+\tilde\Psi_{3^{ijk}} \, , \\%\label{Psi3}
&&\hspace{-8mm} \tilde\Psi_{3^{ijk}} = m_{i}^{p}m_{j}^{m}m_{k}^{q}\,\bigg[X_{pmq}-\frac{1}{D-3}\left( g_{pm}\,X_q-g_{pq}\,X_m\right)\bigg] \, ,
\end{eqnarray}
where $a$ is given by (\ref{typIIa}), $f_p$ can locally be written as (\ref{fpotential}),
\begin{eqnarray}
&&\hspace{-8mm} T_p \equiv \frac{2\,e_{p}}{D-2}\!\left(\frac{\,^{S}\!R}{D-3}+f\right)-\frac{1}{2}f^qe_qf_p+f^q E_{qp}+\frac{2}{D-3}\,X_p\, , \label{defTp}
\end{eqnarray}
${X_q \equiv g^{pm}X_{pmq}}$, and ${X_{pmq}}$ defined in (\ref{Xipj*}) simplifies to
\begin{eqnarray}
&&\hspace{-17mm} X_{pmq} = {\textstyle e_{[q||m]||p}+\frac{1}{2}\!\left(e_{pm}f_q-e_{pq}f_m\right)-\frac{1}{2}\!\left(f_{pm}\,e_q-f_{pq}\,e_m\right)+g_{p[m,u||q]}}\, , \label{defXpmq}
\end{eqnarray}
because for the subtype II(ad) there is
\begin{eqnarray}
&&\hspace{-8mm}
{\textstyle\frac{1}{2} \Big(rf_p\,g_{uu,rr}+g_{uu,rp}-f_{p,u}\Big) = (a_{,p}+f_{p}\,a)\,r +\frac{1}{2}(b_{,p}-f_{p,u})}\, , \label{condIIIa}\\
&&\hspace{7.8mm}
{\textstyle \Big(\frac{1}{2}\,g_{uu,rr}-\frac{1}{4}f^pf_p\Big) = -\frac{1}{D-2}\left(\frac{\,^{S}\!R}{D-3}+f\right)} , \label{condIIIb}\\
&&\hspace{28.8mm}
F_{pq} = 0 \, . \label{condIIIc}
\end{eqnarray}
It can be seen that ${\tilde\Psi_{3^{i}}{}^{_i}{}_{^k}=0}$ which confirms relation ${\Psi_{3^{i}}{}^{_i}{}_{^k}=\Psi_{3T^k}}$, see (\ref{constraints}).

Following the definition in section~2.3 of~\cite{OrtaggioPravdaPravdova:2013} we can thus explicitly write the conditions for the subtypes III(a) and III(b) with triple WAND~${\boldk}$:

\bigskip

\begin{itemize}

\item
The Kundt type~III spacetimes are of \emph{algebraic type}~III(a) if, and only if, ${\Psi_{3T^j}=0}$. Due to (\ref{Psi 3Tj}), which is linear in $r$, this yields the following two conditions that must be satisfied simultaneously:
\begin{eqnarray}
 a_{,p}+f_{p}\,a &=& 0 \, , \label{typIIIa1} \\
 b_{,p}-f_{p,u}  &=& T_p \,. \label{typIIIa2}
\end{eqnarray}
These put specific constraints on the spatial derivatives of the functions $a$ and~$b$ in the expression (\ref{guuprospinttypII}) for $g_{uu}$. Notice that, in view of (\ref{typIIa}) and (\ref{fpotential}), the condition  (\ref{typIIIa1}) can be (locally) integrated as
\begin{equation}
a(u,x)= \frac{1}{4}f^p f_p-\frac{1}{D-2}\left(\frac{\,^{S}\!R}{D-3}+f\right)=\alpha\exp(-{\cal F})\, , \label{aFpot}
\end{equation}
where ${{\cal F}(x,u)}$ is the potential function of~${f_p={\cal F}_{,p}}$ while $\alpha(u)$ is an arbitrary function of the retarded time $u$.

\item
The Kundt type~III spacetimes are of \emph{algebraic type}~III(b) if, and only if, ${\tilde\Psi_{3^{ijk}}=0}$, which is explicitly
\begin{equation}
X_{pmq}=\frac{1}{D-3}\left( g_{pm}\,X_q-g_{pq}\,X_m\right) \, . \label{typIIIb}
\end{equation}
This is the set of constraints on the metric functions ${e_p, f_p, g_{pq}}$. In four spacetime dimensions, ${D=4}$, the condition (\ref{typIIIb}) is satisfied identically, so that III(b)$\equiv$III, while III(a)$\equiv$N. Indeed, in such a case the transverse metric is two-dimensional and thus conformally flat, ${g_{pm}=\Omega\,\delta_{pm}}$, which implies ${X_{pmq}= g_{pm}\,X_q-g_{pq}\,X_m}$ with ${X_q=\Omega^{-1}(X_{22q}+X_{33q})}$  for all combinations of the indices ${p,m,q=2,3}$.

\end{itemize}

\section{Type N Kundt spacetimes}
\label{claasifKundtN}

Kundt spacetime (\ref{obecny Kundt II}) is of \emph{algebraic type}~N with quadruple WAND ${\boldk=\partial_r}$  if, and only if, the four independent conditions (\ref{guuprospinttypII})--(\ref{typIId}) and the three independent conditions (\ref{typIIIa1}), (\ref{typIIIa2}), (\ref{typIIIb}) are all satisfied simultaneously for every combination of the spatial components ${m,n,p,q=2,\ldots,D-1\,}$, i.e., N=III(ab).

Therefore, the only non-vanishing Weyl tensor component is ${\Psi_{4^{ij}}}$ given by (\ref{Psi4D}). For all type~N Kundt geometries, this scalar of the lowest boost weight ${-2}$ is
\begin{equation}
 \Psi_{4^{ij}} =  m_{i}^{p}m_{j}^{q}\, \bigg[\,W_{pq}-\frac{1}{D-2}\, g_{pq}\,W\,\bigg]  \, , \label{Psi4N}
\end{equation}
where the symmetric matrix $W_{pq}$ given by (\ref{WpqD}) simplifies, using (\ref{guuprospinttypII}), (\ref{typIIa}), (\ref{typIId}), (\ref{typIIIa1}), (\ref{typIIIa2}),  to
\begin{eqnarray}
&&\hspace{-25mm} W_{pq} \equiv r\,\bigg[\,\frac{1}{2}\,a\, g_{pq,u}+U_{(p||q)}+U_{(p} f_{q)}\, \bigg] \nonumber\\
&&\hspace{-16mm}
-\frac{1}{4}\Big[\big(c_{,p}-c\,f_{p}\big)_{||q}+\big(c_{,q}-c\,f_{q}\big)_{||p}\Big]
-\frac{1}{2}b\,e_{pq} +\Big(a-\frac{1}{4}f^mf_m\Big)e_{p}e_{q}+Z_{(pq)}\, , \label{WpqN}
\end{eqnarray}
in which $a$ is given by (\ref{typIIa}), ${\,U_p = \frac{1}{2}(f_{p,u}-T_p)-a\,e_p\,}$ is
\begin{eqnarray}
&&\hspace{-24.2mm} U_p \equiv \frac{1}{2}f_{p,u}-\frac{1}{4}f^{q}f_{q}\,e_{p}+\frac{1}{4}f^{q}e_{q}f_{p}-\frac{1}{2}f^{q}E_{qp}-\frac{1}{D-3}\,X_{p}\, , \label{defUp}\\
&&\hspace{-25.4mm} Z_{pq} \equiv
\frac{1}{4}e^{m}e_{m}\,f_{p}f_{q} +e_{p,u||q}-\frac{1}{2}\,g_{pq,uu}-e^{m}E_{mp}f_{q}
+g^{mn}E_{mp}E_{nq}-\frac{2}{D-3}\,X_{p}\,e_{q} \, , \label{Zpq}
\end{eqnarray}
and the trace of ${W_{pq}}$ is ${W=g^{pq}W_{pq}\,}$. The symmetric ${(D-2)\times(D-2)}$ matrix~${\Psi_{4^{ij}}}$ represents the specific amplitudes of Kundt gravitational waves, which in any dimension $D$ are thus always transverse and traceless. This result applies to all Kundt type~N geometries, irrespective of any field equations.

\section{Type O Kundt spacetimes}
\label{claasifKundt0}

Type O Kundt geometries arise when \emph{all} components of the Weyl tensor vanish. From (\ref{Psi4N}) it follows that this occurs for type~N Kundt spacetimes if, and only if,
\begin{equation}
W_{pq}=\frac{1}{D-2}\, g_{pq}\,W  \, . \label{conflat}
\end{equation}
This condition, which is symmetric in the spatial components $p$ and $q$, obviously splits into the part linear in the coordinate~$r$, and the part independent of it.

Due to (\ref{conflat}), the matrix $W_{pq}$ identically vanishes if, and only if, its trace vanishes, ${W_{pq}=0\Leftrightarrow W=0}$. Therefore, there exists a \emph{specific subtype} of conformally flat Kundt geometries, which we may denote as type~O', given by the condition ${W_{pq}=0\,}$. Using (\ref{WpqN}) this is equivalent to
\begin{eqnarray}
&&\hspace{6.4mm} U_{(p||q)}+U_{(p} f_{q)} = -\frac{1}{2}\,a\, g_{pq,u} \ , \label{WpqOa}\\
&&\hspace{-15mm}
\big(c_{,p}-c\,f_{p}\big)_{||q}+\big(c_{,q}-c\,f_{q}\big)_{||p}
=-2b\,e_{pq} +\Big(4a-f^mf_m\Big)e_{p}e_{q}+4Z_{(pq)} \, ,\label{WpqOb}
\end{eqnarray}
where the functions ${a, b, c}$, introduced in (\ref{guuprospinttypII}), must satisfy the constraints (\ref{typIIa}), (\ref{typIIIa1}), (\ref{typIIIa2}), while the functions ${U_p, Z_{pq}}$ were defined in (\ref{defUp}), (\ref{Zpq}).

\section{Type D Kundt spacetimes}
\label{claasifKundtD}

Finally, we may complete the explicit classification of Kundt geometries by identifying those algebraically special spacetimes for which the Weyl scalars of the lowest boost weights ${-2}$ and ${-1}$ vanish.

Let us consider a generic Kundt spacetime of type~II with respect to the double WAND ${\boldk=\partial_r}$, described in section~\ref{claasifKundtI}. If its boost weight ${-2}$ scalar (\ref{Psi4D}) (in which the functions $W_{pq}$ and $W$ are given by (\ref{WpqD}) and (\ref{W}), respectively)  vanishes, ${\Psi_{4^{ij}}=0}$, the vector ${\boldl}$ given by (\ref{Kundt simplest frame}) is WAND and the spacetime is of the \emph{algebraic type}~II$_i$.

If, in addition, all the conditions (\ref{guuprospinttypII})--(\ref{typIId}) are satisfied, the boost weight 0 Weyl scalars also vanish and the spacetime is thus of \emph{algebraic type}~III$_i$ with respect to triple WAND ${\boldk}$ and WAND ${\boldl}$.

Complementary, the Kundt geometries of \emph{algebraic type}~D  are represented by those metrics (\ref{obecny Kundt II}) for which all Weyl scalars of the \emph{two} lowest boost weights, namely $\Psi_{3T^j}$, $\Psi_{3^{ijk}}$ and $\Psi_{4^{ij}}$, are zero. These scalars are explicitly given by expressions (\ref{Psi 3TjD})--(\ref{Psi4D}). All type~D Kundt spacetimes for which \emph{both} ${\boldk}$ \emph{and} ${\boldl}$ given by (\ref{Kundt simplest frame}) \emph{are double WANDs} thus can be written in the form (\ref{obecny Kundt II}) in which the metric functions satisfy the conditions
\begin{eqnarray}
&&\hspace{-10mm} \frac{1}{2} \Big(rf_p\,g_{uu,rr}
+g_{uu,rp}-f_{p,u}\Big)+e_p\Big(\frac{1}{2}\,g_{uu,rr}-\frac{1}{4}f^qf_q\Big)= \nonumber\\
&&\hspace{1.4mm}
r\,\Big(\frac{1}{2}f^q F_{qp}+\frac{1}{D-3}Y_p\Big)
-\frac{1}{4}f^qe_qf_p+\frac{1}{2}f^qE_{qp}+\frac{1}{D-3}\,X_p \ ,\label{Psi 3TjDco}\\
&&\hspace{-10mm} X_{pmq}=\frac{1}{D-3}(\, g_{pm}\,X_q-g_{pq}\,X_m)\ ,\label{Psi3ijkDco1}\\
&&\hspace{-9mm} Y_{pmq} =\frac{1}{D-3}(\, g_{pm}\,Y_q-g_{pq}\,Y_m) \, , \label{Psi3ijkDco2}\\
&&\hspace{-8mm} W_{pq}=\frac{1}{D-2}\, g_{pq}\,W  \, , \label{Psi4Dco}
\end{eqnarray}
where the corresponding functions are defined in (\ref{Xipj*})--(\ref{W}).

Moreover, if the relations (\ref{guuprospinttypII}) with (\ref{typIIa}), (\ref{typIIb}), (\ref{typIIc}), (\ref{typIId}) are also valid, we obtain the particular \emph{subtypes}~D(a),~D(d),~D(c),~D(d), respectively. The subtype D(abcd) is equivalent to (conformally flat) type~O Kundt spacetimes, described in section~\ref{claasifKundt0}.

In the case of \emph{four-dimensional spacetimes} (${D=4}$), the conditions (\ref{Psi3ijkDco1}) and (\ref{Psi3ijkDco2}) are identically satisfied  (due to the arguments analogous to those given at the end of section~\ref{claasifKundtIII}), and the type~D spacetimes are automatically of type~D(bc).

\section{Summary of the results}
\label{summarysub}

Let us summarize the classification scheme of the Kundt spacetimes of type~II with respect to double WAND ${\boldk}$ in an arbitrary dimension $D$. Such algebraically special Kundt geometries (\ref{obecny Kundt II}) are classified into specific distinct subclasses listed in table~\ref{classifKundt}.

\renewcommand{\arraystretch}{1.4}
\begin{table}[ht]
\begin{tabular}{|c||c|c|}
\hline
  type & necessary and sufficient conditions & equations \\
\hline\hline\hline
II(a)  & ${g_{uu}=a(u,x)\,r^2+ b(u,x)\,r+c(u,x)}$  & (\ref{guuprospinttypII}) \\[-1mm]
 & where
${\,a=\frac{1}{4}f^p f_p-\frac{1}{D-2}\big(\frac{\,^{S}\!R}{D-3}+f\big)}$  &  (\ref{typIIa}) \\
\hline
II(b)  & ${\,^{S}\!R_{pq}-\frac{1}{D-2}\,g_{pq}\,^{S}\!R
=-\frac{1}{2}(D-4)\big(f_{pq}-\frac{1}{D-2}\,g_{pq}\,f\big)}$  & (\ref{typIIb}) \\
\hline
II(c)  & ${\,^{S}\!C_{mpnq}=0}$   & (\ref{typIIc}) \\
\hline
II(d)  & ${F_{pq}=0}$  & (\ref{typIId}) \\
\hline\hline\hline
III & II(abcd)   &  \\
\hline
III(a) & ${ a_{,p}+f_{p}\,a = 0\,}$ where $\,a$ is given by (\ref{typIIa})     & (\ref{typIIIa1}), (\ref{typIIa}) \\
       & ${ \ \  b_{,p}-f_{p,u}= T_p}$  &  (\ref{typIIIa2}), (\ref{defTp}) \\
\hline
III(b) & ${X_{pmq}=\frac{1}{D-3}\big( g_{pm}\,X_q-g_{pq}\,X_m\big)}$   & (\ref{typIIIb}), (\ref{defXpmq}) \\
\hline\hline\hline
N & III(ab)   &  \\
\hline\hline\hline
O & N with ${\,W_{pq}=\frac{1}{D-2}\, g_{pq}\,W}$   & (\ref{conflat}), (\ref{WpqN}) \\
\hline
O' & ${\,W_{pq}=0}$   &(\ref{WpqOa}), (\ref{WpqOb}) \\
\hline\hline\hline
D & ${\frac{1}{2} \big(rf_p\,g_{uu,rr}+g_{uu,rp}-f_{p,u}\big)
+e_p\big(\frac{1}{2}\,g_{uu,rr}-\frac{1}{4}f^qf_q\big)\quad}$   &  \\[-1mm]
 & ${=r\,\big(\frac{1}{2}f^q F_{qp}+\frac{1}{D-3}Y_p\big)
-\frac{1}{4}f^qe_qf_p+\frac{1}{2}f^qE_{qp}+\frac{1}{D-3}\,X_p}$   & (\ref{Psi 3TjDco}), (\ref{Xp*}), (\ref{Yp*})  \\
 & ${ X_{pmq}=\frac{1}{D-3}\big(\, g_{pm}\,X_q-g_{pq}\,X_m\big)}$   & (\ref{Psi3ijkDco1}), (\ref{Xipj*}), (\ref{Xp*})  \\
  & ${ Y_{pmq} =\frac{1}{D-3}\big(\, g_{pm}\,Y_q-g_{pq}\,Y_m\big)}$   & (\ref{Psi3ijkDco2}), (\ref{YpmgD}), (\ref{Yp*})  \\
  & ${ W_{pq}=\frac{1}{D-2}\, g_{pq}\,W  }$   & (\ref{Psi4Dco}), (\ref{WpqD}), (\ref{W})  \\
\hline
D(a) & D with II(a)   &  \\
\hline
D(b) & D with II(b)   &  \\
\hline
D(c) & D with II(c)   &  \\
\hline
D(d) & D with II(d)   &  \\
\hline
\end{tabular}
%\vspace{3mm}
\caption{\label{classifKundt} The complete classification scheme of algebraically special Kundt geometries (\ref{obecny Kundt II}) in any dimension $D$ (in the classic ${D=4}$ case, the conditions for II(b), II(c), III(b) are automatically satisfied). The vector ${\boldk=\partial_r}$ is a multiple WAND, and for type~D subclass the vector ${\boldl=\frac{1}{2}g_{uu}\,\mathbf{\partial}_r+\mathbf{\partial}_u}$ is a double WAND.}
\end{table}

In table~\ref{classifKundt} the necessary and sufficient conditions and the corresponding equation references for all the algebraic types and subtypes are given. Recall that $g_{pq}$, where ${p,q=2,3,\ldots,D-1}$, denotes the metric of the transverse Riemannian space, the corresponding Ricci tensor and Ricci scalar are ${\,^{S}\!R_{pq}}$ and ${\,^{S}\!R}$, respectively, and its Weyl tensor is ${\,^{S}\!C_{mpnq}}$. The remaining quantities are defined in (\ref{defkontra})--(\ref{Laplaceguut}).

To illustrate the usefulness of this scheme, in the remaining sections we apply it to three most important subfamilies of the Kundt geometry, namely the pp-waves, the VSI spacetimes, and generalization of the Bertotti--Robinson, Nariai, and Pleba\'{n}ski--Hacyan direct-product spacetimes.

\section{pp-waves}
\label{ppwaves}

The class of pp-waves is defined geometrically by the property that the spacetimes admit a \emph{covariantly constant null vector field} $\boldk$, see~\cite{Stephani:2003,GriffithsPodolsky:2009}. All the optical scalars (\ref{optical scalars}) thus vanish, so that the pp-waves necessarily belong to the Kundt family (\ref{obecny Kundtuv prostorocas}) with ${\boldk=\partial_r}$. In view of (\ref{Christ1}), (\ref{Christ2}), the defining condition ${0=k_{a;b}=\frac{1}{2}g_{ab,r}}$ immediately implies that for pp-waves all the metric functions must be independent of the coordinate $r$, i.e., the line element can be written in the Brinkmann form~\cite{Bri25} as
\begin{equation}
\dd s^2 = g_{pq}\,\dd x^p\dd x^q+2\,e_p\,\dd u\,\dd x^p -2\,\dd u\,\dd r+c\,\dd u^2 \, , \label{pp}
\end{equation}
where ${\,g_{pq}(u,x),\, e_p(u,x),\, c(u,x)\,}$ are function of $u$ and ${x\equiv(x^2, x^3, \ldots, x^{D-1})}$ only. This is the particular case of the metric (\ref{obecny Kundt II}), (\ref{guuprospinttypII}) in the case when
\begin{equation}
f_p=0\, ,\qquad  a=0=b\, . \label{ppspecialforms}
\end{equation}
All pp-waves are thus algebraically special, that is, of type~II(d) or more special with the multiple WAND ${\boldk}$, in agreement with \cite{Coley:2008,OrtaggioPravdaPravdova:2013}. The relevant quantities which enter in table~\ref{classifKundt} are
\begin{eqnarray}
&& f = 0 \, ,\quad f_{pq}=0\, ,\quad  F_{pq}=0\, ,\label{ppc}\\
&& Y_{pmq} =0 = Y_q\, ,\quad g_{uu,rr}=0=g_{uu,rp} \, ,\label{ppY}\\
&& T_p = \frac{2}{D-3}\Big(X_p+\frac{1}{D-2}{\,^{S}\!R}\,e_{p}\Big), \label{ppTp}\\
&& E_{pq} = e_{[p||q]}\,+{\textstyle \frac{1}{2}}g_{pq,u}\, , \label{ppk}\\
&& X_{pmq}= e_{[q||m]||p}+g_{p[m,u||q]} \, ,\label{ppe}\\
&& W_{pq}=-{\textstyle \frac{1}{2}\,c_{\,||p||q}+ e_{(p,u||q)}-\frac{1}{2}\,g_{pq,uu} +g^{mn}E_{mp}E_{nq} }\, ,\label{ppg}
\end{eqnarray}
and their contractions read
\begin{eqnarray}
&& X_q= g^{mn}\big(e_{[q||m]||n}+g_{m[n,u||q]}\big)\, ,\label{ppf}\\
&& W=-{\textstyle \frac{1}{2}}\,\triangle\, c+g^{pq}\big( e_{(p,u||q)}-{\textstyle \frac{1}{2}}\,g_{pq,uu}\big) +g^{mn}g^{pq}E_{mp}E_{nq} \, .\label{pph}
\end{eqnarray}
We thus obtain the complete algebraic classification of all pp-wave geometries, as summarized in table~\ref{classifpp}.
\begin{table}[ht]
\begin{tabular}{|c||l|}
\hline
  type & \hskip20mm necessary and sufficient conditions  \\
\hline\hline\hline
II(a)  & ${{\,^{S}\!R}=0}$   \\
\hline
II(b)  & ${\,^{S}\!R_{pq}=\frac{1}{D-2}\,g_{pq}\,^{S}\!R }$  \\
\hline
II(c)  & ${\,^{S}\!C_{mpnq}=0}$  \\
\hline
II(d)  & \ always   \\
\hline\hline\hline
III & ${g_{pq}=\delta_{pq}}$    \\
\hline
III(a) & ${g_{pq}=\delta_{pq}}$ \quad and\quad ${ X_p = 0}$   \\
\hline
III(b) & ${g_{pq}=\delta_{pq}}$ \quad and\quad
${X_{pmq}=\frac{1}{D-3}\big( \delta_{pm}\,X_q-\delta_{pq}\,X_m\big)}$  \\
\hline\hline\hline
N & ${g_{pq}=\delta_{pq}}$ \quad and\quad ${X_p=0=X_{pmq}}$    \\
\hline\hline\hline
O & ${g_{pq}=\delta_{pq}}$ \quad and\quad ${X_p=0=X_{pmq}}$ \quad and\quad ${\,W_{pq}=\frac{1}{D-2}\, \delta_{pq}\,W}$   \\
\hline
O' & ${g_{pq}=\delta_{pq}}$ \quad and\quad ${X_p=0=X_{pmq}}$ \quad and\quad ${\,W_{pq}=0=W}$    \\
\hline\hline\hline
D & ${X_p=0=X_{pmq}}$ \quad and\quad  ${ W_{pq}=\frac{1}{D-2}\, g_{pq}\,W  }$   \\
\hline
D(a) & ${X_p=0=X_{pmq}}$ \quad and\quad  ${ W_{pq}=\frac{1}{D-2}\, g_{pq}\,W  }$ \quad and\quad
  ${{\,^{S}\!R}=0}$    \\
\hline
D(b) & ${X_p=0=X_{pmq}}$ \quad and\quad   ${ W_{pq}=\frac{1}{D-2}\, g_{pq}\,W  }$ \quad and\quad
  ${\,^{S}\!R_{pq}=\frac{1}{D-2}\,g_{pq}\,^{S}\!R }$    \\
\hline
D(c) & ${X_p=0=X_{pmq}}$ \quad and\quad ${ W_{pq}=\frac{1}{D-2}\, g_{pq}\,W  }$ \quad and\quad
  ${\,^{S}\!C_{mpnq}=0}$   \\
\hline
D(d) & identical to D    \\
\hline
\end{tabular}
%\vspace{3mm}
\caption{\label{classifpp} The complete classification scheme of all pp-wave geometries (\ref{pp})
in~any dimension $D$.}
\end{table}

Notice that the classification into the subtypes II(a), II(b), II(c) is directly determined by vanishing of the three distinct components of the unique geometric decomposition of the Riemann curvature tensor ${\,^{S}\!R_{mpnq}}$ of the  ${(D-2)}$-dimensional transverse space, namely the Ricci scalar ${\,^{S}\!R}$, the traceless part of the Ricci tensor ${\,^{S}\!R_{pq}-\frac{1}{D-2}\,g_{pq}\,^{S}\!R}$, and the Weyl tensor ${\,^{S}\!C_{mpnq}}$, respectively.

Consequently, the pp-waves are of type III$\equiv$II(abc)=II(abcd) and more special types N and O if, and only if, their \emph{transverse space is flat}, ${g_{pq}=\delta_{pq}}$, i.e., ${\,^{S}\!R_{mpnq}=0}$. Further subclassification of the type~III, N and O pp-waves is then given by mutual relations and possible vanishing of the functions $X_{pmq},X_p$ and $W_{pq}, W$, as given by expressions (\ref{ppe})--(\ref{pph}). Due to ${g_{pq}=\delta_{pq}}$, these simplify to
\begin{eqnarray}
X_{pmq}&=& e_{[q,m],p} \, ,\label{ppes}\\
X_q    &=& \delta^{mn}e_{[q,m],n}\, ,\label{ppfs}\\
W_{pq} &=&-{\textstyle \frac{1}{2}\,c_{,pq}+ e_{(p,uq)}+\delta^{mn}e_{[m,p]}\,e_{[n,q]} }\, ,\label{ppgs}\\
W      &=&-{\textstyle \frac{1}{2}}\,\triangle\, c
+g^{pq}e_{(p,uq)}+\delta^{mn}\delta^{pq}e_{[m,p]}\,e_{[n,q]} \, .\label{pphs}
\end{eqnarray}
In particular, when ${e_p=0}$ (in the absence of a gyratonic matter) this further simplifies to ${X_{pmq}=0=X_q}$, ${W_{pq}=-{\textstyle \frac{1}{2}\,c_{,pq}}\,}$, and ${\,W=-{\textstyle \frac{1}{2}}\,\triangle\, c\,}$. Therefore, the class of pp-wave metrics
\begin{equation}
\dd s^2 = \delta_{pq}\,\dd x^p\dd x^q-2\,\dd u\,\dd r+c\,\dd u^2 \, , \label{ppN}
\end{equation}
is necessarily of~type~N, unless ${{\,c_{,pq}}=\frac{1}{D-2}\, \delta_{pq}\,\triangle\, c\,}$, in which case the spacetimes are type O, cf. \cite{ColFusHerPel06}.

For pp-wave geometries of type D, the transverse space with metric $g_{pq}$ can not be flat (otherwise, they would become of type~O). In such a case, the general expressions (\ref{ppe})--(\ref{pph}) must be used.

Recall that the conditions for types II(b), II(c), III(b) are identically satisfied in the ${D=4}$ subcase.

Our results extend those presented in Proposition 7.3 of \cite{OrtaggioPravdaPravdova:2013}.

\section{VSI spacetimes}
\label{VSIspacetimes}

The VSI spacetimes have the property that their \emph{scalar curvature invariants of all orders vanish} identically. They necessarily belong to  the Kundt class and, relative to~${\boldk}$, their Riemann tensor is of type III or more special (Proposition 5.2 of \cite{OrtaggioPravdaPravdova:2013}).
As shown in \cite{ColFusHerPel06}, see equations (20)--(23) therein, these spacetimes must be of the form (\ref{obecny Kundt II}) with $g_{uu}$ quadratic in $r$ and flat transverse space ${g_{pq}=\delta_{pq}}$,
\begin{equation}
\hskip-15mm
%\dd s^2 = \delta_{ij} \,\dd x^i\dd x^j+2\,(e_i+ f_i \,r)\,\dd u\,\dd x^i -2\,\dd u\,\dd r+(a\,r^2+b\,r+c)\,\dd u^2
\dd s^2 = \delta_{pq} \,\dd x^p\dd x^q+2\,(e_p+ f_p \,r)\,\dd u\,\dd x^p -2\,\dd u\,\dd r+(a\,r^2+b\,r+c)\,\dd u^2
\, , \label{VSI}
\end{equation}
where
\begin{eqnarray}
&&\hskip2.8mm a={\textstyle\frac{1}{4}}f^p f_p\,,\qquad f_{pq}=0\,,\qquad f=0\,,\qquad F_{pq}=0\,, \label{VSI1}\\
&&\,^{S}\!R=0\,,\qquad \,^{S}\!R_{pq}=0\,,\qquad \,^{S}\!C_{mpnq}=0\,. \label{VSI2}
\end{eqnarray}
In full agreement with~\cite{ColFusHerPel06,OrtaggioPravdaPravdova:2013}, all such spacetimes are of Weyl type~III or more special, since the conditions in table~\ref{classifKundt} for II(a), II(b), II(c) and II(d) are automatically satisfied.

The possible subtypes are determined by the conditions listed in table~\ref{classifVSI}, in which
\begin{eqnarray}
&&\hspace{-14mm} T_p =-\frac{1}{2}f^qe_qf_p+f^q  e_{[q,p]}+\frac{2}{D-3}\,X_p\, , \label{VSITp}\\
&&\hspace{-14mm} U_p = \frac{1}{2}(f_{p,u}-T_p)-\frac{1}{4}f^q f_q\,e_p\, , \label{defUpVSI}\\
&&\hspace{-14mm} Z_{pq} =
\frac{1}{4}e^{m}e_{m}\,f_{p}f_{q} +e_{p,uq}-e^{m} e_{[m,p]}f_{q}
+g^{mn} e_{[m,p]} e_{[n,q]}-\frac{2}{D-3}\,X_{p}\,e_{q} \, , \label{ZpqVSI}\\
&&\hspace{-14mm}  X_{pmq}={\textstyle e_{[q,m],p}+\frac{1}{2}\!\left( e_{(p,m)}f_q- e_{(p,q)}f_m\right)}
\, ,\label{VSIe}\\
&&\hspace{-14mm} W_{pq} = r\,\big[\,U_{(p,q)}+U_{(p} f_{q)}\, \big]\nonumber\\
&&\hspace{-4mm} -{\textstyle
\frac{1}{4}\big[\big(c_{,p}-c\,f_{p}\big)_{,q}+\big(c_{,q}-c\,f_{q}\big)_{,p}\big]
-\frac{1}{2}b\, e_{(p,q)} +Z_{(pq)}}\, ,\label{VSIg}
\end{eqnarray}
and
\begin{eqnarray}
&&\hspace{-10mm} X_q={\textstyle \delta^{mn}e_{[q,m],n}+\frac{1}{2}\delta^{mn} e_{m,n}f_q-\frac{1}{2} f^pe_{(p,q)}}\, ,\label{VSIf}\\
&&\hspace{-10mm} W=r\,\delta^{pq}\big(U_{p,q}+U_{p} f_{q}\, \big)-{\textstyle
\frac{1}{2}\delta^{pq}\big(c_{,p}-c\,f_{p}\big)_{,q}
-\frac{1}{2}b\, \delta^{pq}e_{p,q} + \delta^{pq}Z_{pq}} \, .\label{VSIh}
\end{eqnarray}
\begin{table}[ht]
\begin{tabular}{|c||c|}
\hline
  type & \hskip0mm necessary and sufficient conditions  \\
\hline\hline\hline
\hline
III(a) & ${ a_{,p}+f_{p}\,a = 0\,}$ where ${\,a={\textstyle\frac{1}{4}}f^p f_p}$        \\
       & ${\hskip2.1mm b_{,p}-f_{p,u}= T_p}$   \\
\hline
III(b) & ${X_{pmq}=\frac{1}{D-3}\big( \delta_{pm}\,X_q-\delta_{pq}\,X_m\big)}$  \\
\hline\hline\hline
N &  III(ab)   \\
\hline\hline\hline
O & N with  ${\,W_{pq}=\frac{1}{D-2}\, \delta_{pq}\,W}$   \\
\hline
O' &  ${\,W_{pq}=0=W}$    \\
\hline
\end{tabular}
%\vspace{3mm}
\caption{\label{classifVSI} The complete classification scheme of all VSI geometries (\ref{VSI})
in~any dimension $D$.}
\end{table}

This agrees with the results presented in~\cite{ColFusHerPel06}, and extends them because no field equations and gauge fixing have been employed.

\section{Generalized Bertotti--Robinson, Nariai, and other type D and O spacetimes}
\label{DObackgrounds}

Finally, let us consider a subclass of the Kundt geometries of the form
\begin{equation}
\dd s^2 = g_{pq} \,\dd x^p\dd x^q -2\,\dd u\,\dd r+a\,r^2\dd u^2 \, , \label{Bertmetric}
\end{equation}
which is the special case of (\ref{obecny Kundt II}) with
\begin{equation}
e_p=0\,,\qquad f_p=0\,,\qquad b=0\,,\qquad c=0\,. \label{Bertcoef}
\end{equation}
As recently dicussed in~\cite{KrtousPodolskyZelnikovKadlecova:2012}, such spacetimes include interesting CSI backgrounds on which the Kundt waves and gyratons in any (higher) dimension propagate.

Algebraic classification of these metrics depends on the following functions:
\begin{eqnarray}
&&\hskip-10mm f = 0 \, ,\quad f_{pq}=0\, ,\quad  F_{pq}=0\, ,\label{dpc}\\
&&\hskip-10mm
 Y_{pmq} =0 = Y_q\,, \label{dpy} \\
 &&\hskip-10mm
g_{uu,rp}=2a_{,p}\,r \,, \label{dpz} \\
&&\hskip-10mm
 T_p = \frac{2}{D-3}\,X_p\,, \label{dpTp}\\
&&\hskip-10mm
 X_{pmq}= g_{p[m,u||q]} \, ,\label{dpe}\\
&&\hskip-10mm
 X_q= g^{mn}g_{m[n,u||q]}\, ,\label{dpf}\\
&&\hskip-10mm
 W_{pq}=- {\textstyle \frac{1}{2}\,a_{||p||q}\,r^2 +\frac{1}{2}a\,r\,g_{pq,u}
-\frac{1}{2}g_{pq,uu}+\frac{1}{4}g^{mn}g_{mp,u}\,g_{nq,u} }\, , \label{dpfW}\\
&&\hskip-10mm
W=-{\textstyle \frac{1}{2}\,\triangle\, a\,r^2
+\frac{1}{2}a\,r\,g^{pq}g_{pq,u}-\frac{1}{2}g^{pq}g_{pq,uu}
+\frac{1}{4}g^{mn}g^{pq}g_{mp,u}\,g_{nq,u}}\, .\label{dph}
\end{eqnarray}
The corresponding subtypes, as determined generally in table~\ref{classifKundt}, can thus be summarized in table~\ref{classifKundtdp}.
Notice that the condition for subtype~O' splits, using (\ref{dpfW}), into
${\,a\,g_{pq,u}=0 }$ and ${g_{pq,uu}=\frac{1}{2}g^{mn}g_{mp,u}\,g_{nq,u} }$. There are thus two distinct subclasses, namely ${a=0}$ and ${g_{pq,u}=0}$.

\begin{table}[ht]
\begin{tabular}{|c||c|}
\hline
  type & necessary and sufficient conditions \\
\hline\hline\hline
II(a)   &  ${\,a=-\frac{1}{(D-2)(D-3)}\,^{S}\!R}$  \\
\hline
II(b)  & ${\,^{S}\!R_{pq}=\frac{1}{D-2}\,g_{pq}\,^{S}\!R}$   \\
\hline
II(c)  & ${\,^{S}\!C_{mpnq}=0}$  \\
\hline
II(d)  & always \\
\hline\hline\hline
III & II(abcd)    \\
\hline
III(a) & ${ \,^{S}\!R_{,p}= 0}$\,,\quad ${ X_q= g^{mn}g_{m[n,u||q]}=0}$  \\
\hline
III(b) & ${X_{pmq}=\frac{1}{D-3}\big( g_{pm}\,X_q-g_{pq}\,X_m\big)}$  \\
\hline\hline\hline
N & III(ab)    \\
\hline\hline\hline
O & N with ${\,W_{pq}=\frac{1}{D-2}\, g_{pq}\,W}$    \\
\hline
O' & ${\,W_{pq}=0}$   \\
\hline\hline\hline
D & ${ a_{,p}= 0}$    \\[-1mm]
  & ${ X_q=0 \Leftrightarrow   X_{pmq}=0}$    \\[-1mm]
  & ${ W_{pq}=\frac{1}{D-2}\, g_{pq}\,W  }$   \\
\hline
D(a) & D with II(a)   \\
\hline
D(b) & D with II(b)   \\
\hline
D(c) & D with II(c)   \\
\hline
D(d) & D with II(d)   \\
\hline
\end{tabular}
%\vspace{3mm}
\caption{\label{classifKundtdp} The complete classification scheme of all Kundt geometries (\ref{Bertmetric}) in any dimension $D$ with respect to multiple WAND ${\boldk=\partial_r}$ and (possibly) double WAND ${\boldl=\frac{1}{2}ar^2\mathbf{\partial}_r+\mathbf{\partial}_u}$.}
\end{table}

\begin{table}[ht]
\begin{tabular}{|c||c|}
\hline
  type & necessary and sufficient conditions \\
\hline\hline\hline
D=D(d) & always   \\
\hline
D(a) & ${\,a=-\frac{1}{(D-2)(D-3)}\,^{S}\!R=\hbox{const.}}$    \\
\hline
D(b) & ${\,^{S}\!R_{pq}=\frac{1}{D-2}\,g_{pq}\,^{S}\!R}$  \\
\hline
D(c) & ${\,^{S}\!C_{mpnq}=0}$   \\
\hline\hline\hline
O & D(abcd)= D(abc)   \\
\hline
\end{tabular}
%\vspace{3mm}
\caption{\label{classifKundtdpspec} Algebraic classification of the direct-product Kundt geometries (\ref{Bertmetricspec}) in any dimension $D$. These involve generalizations of the Bertotti--Robinson, (anti-) Nariai, and Pleba\'{n}ski--Hacyan spacetimes of type~D or~O.}
\end{table}

An important subfamily of (\ref{Bertmetric}) arises when $a$ is \emph{constant} and  ${g_{pq,u}=0\,}$:
\begin{equation}
\dd s^2 = g_{pq}(x) \,\dd x^p\dd x^q -2\,\dd u\,\dd r+a\,r^2\dd u^2 \, . \label{Bertmetricspec}
\end{equation}
In such a case, ${X_{pmq}=0=X_q}$ and ${W_{pq}=0=W}$, so that the spacetimes can only be of algebraic type D or type O, as described in table~\ref{classifKundtdpspec}. In fact, these are the \emph{direct-product spacetimes}, where the first part is a ${(D-2)}$-dimensional Riemannian space with metric $g_{pq}$, while the second part is a 2-dimensional Lorentzian spacetime which has a constant curvature. By performing the transformation
\begin{equation}\label{UVtrans}
    U = \frac1{a\, u}\;,\quad V = \frac4{a\, r} + 2\,u \, ,
\end{equation}
the metric (\ref{Bertmetricspec}) is put to the canonical form
\begin{equation}\label{tempmtrcan}
\dd s^2 = g_{pq}(x) \,\dd x^p\dd x^q - \frac{2\,\dd U\dd V}{ \big(1-{\textstyle\frac12}a\,UV\big)^2}  \, .
\end{equation}
According to the sign of the constant ${a}$ there are three possibilities: for ${a=0}$ the temporal surface ${x=}$const. is the 2-dimensional Minkowski space ${M_2}$, for ${a>0}$ it is the \mbox{2-dimensional} de~Sitter space ${dS_2}$, and for ${a<0}$ we get the 2-dimensional anti-de~Sitter space ${AdS_2}$. The scale of the (anti-)de~Sitter space is given by ${\ell=1/\sqrt{|a|}}$.

For generic type~D direct-product spacetimes (\ref{Bertmetricspec}), or equivalently (\ref{tempmtrcan}), the transverse metric $g_{pq}$ \emph{need not} be of constant curvature. However, for the subtype D(a), the Gaussian curvature $a$ is \emph{uniquely related} to the \emph{constant Ricci scalar} ${\,^{S}\!R}$ of the transverse $(D-2)$-dimensional Riemannian space. The metrics then represent higher-dimensional generalizations of the well-known ${D=4}$ Bertotti--Robinson, \mbox{(anti-)Nariai}, and Pleba\'{n}ski--Hacyan spacetimes of types~D or~O, for which the two-dimensional transverse Riemannian spaces are either $S^2$, $E^2$, or $H^2$ of \emph{constant} positive, zero, or negative curvatures, see chapter~7 of~\cite{GriffithsPodolsky:2009} for more details.

General conditions for direct-product spacetimes to be of type~O were worked out a long time ago in the pioneering work \cite{Ficken:1939}, and the fact that such spacetimes must be of type~D(d) or~O was noticed in \cite{PraPraOrt07}, see Proposition 4 therein. In our recent work \cite{KrtousPodolskyZelnikovKadlecova:2012} we analyzed in detail exact type~II Kundt gravitational waves and gyratons propagating on such direct-product backgrounds.

\section{Conclusions}
\label{conclusions}

We derived the complete and explicit algebraic classification, based on the null alignment properties of the Weyl tensor, of the general Kundt class of spacetimes (\ref{obecny Kundtuv prostorocas}) in an arbitrary dimension~$D$ for the case when the generator ${\boldk}$ of optically privileged null congruence with vanishing expansion, twist and shear is a (multiple) WAND. The results apply to any metric theory of gravity which admits the Kundt geometries. Classification of the Kundt spacetimes in standard general relativity is obtained by setting ${D=4}$ (and applying the Einstein field equations).

We calculated all components of the Riemann and Ricci curvature tensors, see (\ref{Riemann rprq - Kundt})--(\ref{Ricci uu - Kundt}), and projecting the corresponding Weyl tensor onto the natural null frame (\ref{Kundt simplest frame}) we showed that the Kundt geometries are of type~I(b) (or more special) with ${\boldk=\partial_r}$ being~WAND aligned with the optically privileged null congruence.

The spacetimes become algebraically special with respect to multiple WAND ${\boldk}$ if, and only if, the metric functions $g_{up}$ are (at most) linear in the affine parameter $r$ along the privileged null congruence with vanishing optical scalars, see the line element (\ref{obecny Kundt II}). For such a generic metric we explicitly evaluated all (rather complicated) coordinate components of the Riemann, Ricci and Weyl tensors, as summarized in~\ref{appendixB}. The corresponding Weyl scalars (of various boost weights and spins) with respect to the most natural null frame (\ref{Kundt simplest frame}) have surprisingly simple structure (\ref{Psi2s})--(\ref{Psi4D}). This enabled us to determine all possible algebraic types with respect to the multiple WAND~${\boldk}$ (and potential additional WAND ${\boldl}$), including the refinement to various subtypes, namely II(a), II(b), II(c), II(d), III(a), III(b), N, O, II$_i$, III$_i$, D(a), D(b), D(c), and D(d), see sections~\ref{claasifKundtI}--\ref{claasifKundtD} respectively. The explicit conditions, given in table~\ref{classifKundt} of section~\ref{summarysub}, are expressed in an invariant geometric form. Some of them are identically satisfied in the ${D=4}$ case.

In the final sections~\ref{ppwaves}--\ref{DObackgrounds} we applied our classification scheme to the most important subclasses of the Kundt geometries, namely the pp-waves, the VSI spacetimes, and the generalization of direct-product spacetimes of any dimension. The main results are contained in tables~\ref{classifpp}--\ref{classifKundtdpspec}.

We hope that these results will be useful for further studies of the Kundt family of geometries and its various interesting representatives. For example, in a subsequent paper~\cite{PodolskySvarc:2013} extending the preliminary analysis in~\cite{PodolskySvarc:2012b}, we will relate the algebraic classification of these spacetimes to specific local motion of test particles, as given by the deviation of geodesics. This will further clarify their geometrical properties and physical interpretation.

\section*{Acknowledgements}

We are grateful to Marcello Ortaggio, Vojta Pravda, Alena Pravdov\'a and an anonymous referee for their very useful comments on the manuscript. J.P.~was supported by the grant GA\v{C}R P203/12/0118 and R.\v{S}. by the grant GA\v{C}R~202/09/0772.

\newpage
\appendix

\section{Curvature tensors for a general Kundt geometry}
\label{appendixA}

For the most general Kundt line element (\ref{obecny Kundtuv prostorocas}) the Christoffel symbols read
\begin{eqnarray}
\Gamma^r_{ru} \rovno -\textstyle{\frac{1}{2}}g_{uu,r}+\frac{1}{2}g^{rp}g_{up,r} \, , \\
\Gamma^r_{rp} \rovno -\textstyle{\frac{1}{2}}g_{up,r} \, , \\
\Gamma^r_{uu} \rovno \textstyle{\frac{1}{2}}\,\big[\!-\!g^{rr}g_{uu,r}-g_{uu,u}+g^{rp}(2g_{up,u}-g_{uu,p})\big] , \\
\Gamma^r_{uq} \rovno \textstyle{\frac{1}{2}}\left[-g^{rr}g_{uq,r}-g_{uu,q}+g^{rp}(2g_{p(u,q)}-g_{uq,p})\right] , \\
\Gamma^r_{pq} \rovno \textstyle{\frac{1}{2}}\left[\,g_{pq,u}-2g_{u(p,q)}+g^{rm}(2g_{m(p,q)}-g_{pq,m})\right] , \\
\Gamma^p_{ru} \rovno \textstyle{\frac{1}{2}}g^{pq}g_{uq,r} \ , \\
\Gamma^p_{uu} \rovno \textstyle{\frac{1}{2}}\,\big[\!-\!g^{rp}g_{uu,r}+g^{pq}(2g_{uq,u}-g_{uu,q})\big] , \\
\Gamma^p_{um} \rovno \textstyle{\frac{1}{2}}\left[-g^{rp}g_{um,r}+g^{pq}(2g_{q(u,m)}-g_{um,q})\right] , \\
\Gamma^m_{pq} \rovno \textstyle{\frac{1}{2}}g^{mn}(2g_{n(p,q)}-g_{pq,n}) \, , \\
\Gamma^u_{uu} \rovno \textstyle{\frac{1}{2}}g_{uu,r} \, ,\label{Christ1} \\
\Gamma^u_{up} \rovno \textstyle{\frac{1}{2}}g_{up,r} \, ,\label{Christ2}
\end{eqnarray}
where the contravariant metric coefficients of (\ref{obecny Kundtuv prostorocas}) are
\begin{equation}
g^{rp}= g^{pq}g_{uq}\,,\qquad g^{rr}=-g_{uu}+g^{pq}g_{up}\,g_{uq}\,,
\label{cocontrarelations}
\end{equation}
with $g^{pq}$ denoting the inverse of $g_{pq}$ (inversely, ${g_{up}= g_{pq}\,g^{rq}}$, ${g_{uu}=-g^{rr}+g_{pq}\,g^{rp}g^{rq}}$).
A straightforward but lengthy calculation gives the following coordinate components of the Riemann curvature tensor in a fully explicit and  compact form
\begin{eqnarray}
&&\hspace{-13mm}
R_{rprq} = 0\, , \label{Riemann rprq - Kundt} \\
&&\hspace{-13mm}
R_{rpru} = -\textstyle{\frac{1}{2}}g_{up,rr} \, , \label{Riemann rpru - Kundt} \\
&&\hspace{-13mm}
R_{ruru} = -\textstyle{\frac{1}{2}}g_{uu,rr}+\textstyle{\frac{1}{4}}g^{pq}g_{up,r}g_{uq,r} \, , \label{Riemann ruru - Kundt} \\
&&\hspace{-13mm}
R_{rpmq} = 0 \, , \label{Riemann rpkq - Kundt} \\
&&\hspace{-13mm}
R_{rpuq} = \textstyle{\frac{1}{2}}g_{up,rq}+\textstyle{\frac{1}{4}}g_{up,r}g_{uq,r}-\textstyle{\frac{1}{4}}g^{mn}g_{um,r}\left(2g_{n(p,q)}-g_{pq,n}\right) , \\
&&\hspace{-13mm}
R_{rupq} = g_{u[p,q],r} \, , \\
&&\hspace{-13mm}
R_{ruup} = g_{u[u,p],r}+\textstyle{\frac{1}{4}}g^{rq}g_{up,r}g_{uq,r}-\textstyle{\frac{1}{4}}g^{mn}g_{um,r}\left(2g_{n(u,p)}-g_{up,n}\right) , \\
&&\hspace{-13mm}
R_{mpnq} = \,^{S}\!R_{mpnq} \, , \\
&&\hspace{-13mm}
R_{upmq} = g_{p[m,q],u}-g_{u[m,q],p} \nonumber \\
&&\ +\textstyle{\frac{1}{4}}\left[g_{um,r}\left(g_{pq,u}-2g_{u(p,q)}\right)-g_{uq,r}\left(g_{pm,u}-2g_{u(p,m)}\right)\right] \nonumber \\
&&\ +\textstyle{\frac{1}{4}}g^{rn}\left[g_{um,r}\left(2g_{n(p,q)}-g_{pq,n}\right)-g_{uq,r}\left(2g_{n(p,m)}-g_{pm,n}\right)\right] \nonumber \\
&&\ +\textstyle{\frac{1}{4}}g^{ns}\left(2g_{s(u,q)}-g_{uq,s}\right)\left(2g_{n(p,m)}-g_{pm,n}\right) \nonumber \\
&&\ -\textstyle{\frac{1}{4}}g^{ns}\left(2g_{s(u,m)}-g_{um,s}\right)\left(2g_{n(p,q)}-g_{pq,n}\right) , \\
&&\hspace{-13mm}
R_{upuq} = g_{u(p,q),u}-\textstyle{\frac{1}{2}}\left(g_{pq,uu}+g_{uu,pq}\right) +\textstyle{\frac{1}{4}}g^{rr}g_{up,r}g_{uq,r}\nonumber \\
&&\ -\textstyle{\frac{1}{4}}g_{uu,r}\left[2g_{u(p,q)}-g_{pq,u}-g^{rm}\left(2g_{m(p,q)}-g_{pq,m}\right)\right] \nonumber \\
&&\ +\textstyle{\frac{1}{4}}g_{up,r}\left[g_{uu,q}-g^{rm}\left(2g_{m(u,q)}-g_{uq,m}\right) \right] \nonumber \\
&&\ +\textstyle{\frac{1}{4}}g_{uq,r}\left[g_{uu,p}-g^{rm}\left(2g_{m(u,p)}-g_{up,m}\right)\right] \nonumber \\
&&\ +\textstyle{\frac{1}{4}}g^{mn}\left(2g_{n(u,p)}-g_{up,n}\right)\left(2g_{m(u,q)}-g_{uq,m}\right) \nonumber \\
&&\ -\textstyle{\frac{1}{4}}g^{mn}\left(2g_{un,u}-g_{uu,n}\right)\left(2g_{m(p,q)}-g_{pq,m}\right) , \label{Riemann upuq - Kundt}
\end{eqnarray}
where $p,q,m,n,s$ denote the spatial components (and derivatives with respect to)~$x$. The superscript ``${\,^{S}\,}$'' labels tensor quantities corresponding to the spatial metric $g_{pq}$, with derivatives taken only with respect to the coordinates~$x$. The corresponding coordinate components of the Ricci tensor are\footnote{Let us remark that the first term in $R_{us}$ is missing in the corresponding component (25) presented in~\cite{PodolskyZofka:2009}. Other components are equivalent using the relations
$(\,\ln\sqrt{g})_{,u}=\Gamma^a_{ua}=\frac{1}{2}g^{pq}(2g_{q(u,p)}-g_{up,q})$ and
$(\,\ln\sqrt{g})_{,p}=\Gamma^a_{pa}=\Gamma^q_{pq}=\frac{1}{2}g^{qm}(2g_{m(p,q)}-g_{pq,m})$,
where
${g\equiv \det g_{pq}=-\det g_{ab}}$.}
\begin{eqnarray}
&&\hspace{-10mm}
R_{rr}= 0\, , \label{Ricci rr - Kundt}\\
&&\hspace{-10mm}
R_{rp}= -\textstyle{\frac{1}{2}}g_{up,rr} \, ,\label{Ricci rk - Kundt}\\
&&\hspace{-10mm}
R_{ru}= -\textstyle{\frac{1}{2}}g_{uu,rr}+\textstyle{\frac{1}{2}}g^{rp}g_{up,rr}+\textstyle{\frac{1}{2}}g^{pq}g_{up,rq} \nonumber \\
&& +\textstyle{\frac{1}{2}}g^{pq}g_{up,r}g_{uq,r}-\textstyle{\frac{1}{4}}g^{pq}g^{mn}g_{um,r}(2g_{np,q}-g_{pq,n}) \, , \label{Ricci ru - Kundt} \\
&&\hspace{-10mm}
R_{pq}= \!^{S}\!R_{pq}-g_{u(p,q),r}-\textstyle{\frac{1}{2}}g_{up,r}g_{uq,r}+\textstyle{\frac{1}{2}}g^{mn}g_{um,r}(2g_{n(p,q)}-g_{pq,n})\, ,
\label{Ricci pq - Kundt}\\
&&\hspace{-10mm}
R_{us}= -\textstyle{\frac{1}{2}}g^{rr}g_{us,rr}-g_{u[u,s],r}+g^{rp}\left(\textstyle{\frac{1}{2}}g_{up,sr}-g_{us,pr}\right) \nonumber \\
&& +g^{pq}(g_{p[s,q],u}-g_{u[s,q],p})-\textstyle{\frac{1}{2}}g^{rp}g_{us,r}g_{up,r} \nonumber \\
&& +\textstyle{\frac{1}{4}}g^{pq}g^{rm}\left[4g_{uq,r}g_{s[p,m]}+g_{us,r}(2g_{m(p,q)}-g_{pq,m})\right] \nonumber \\
&& +\textstyle{\frac{1}{4}}g^{pq}\big[2g_{up,r}g_{uq,s}-g_{us,r}(2g_{up,q}-g_{pq,u})\big] \nonumber \\
&& +\textstyle{\frac{1}{4}}g^{pq}g^{mn}\left(2g_{n(u,q)}-g_{uq,n}\right)\left(2g_{m(p,s)}-g_{ps,m}\right) \nonumber \\
&& -\textstyle{\frac{1}{4}}g^{pq}g^{mn}\left(2g_{n(u,s)}-g_{us,n}\right)\left(2g_{mp,q}-g_{pq,m}\right) , \label{Ricci uk - Kundt}\\
&&\hspace{-10mm}
R_{uu}= -\textstyle{\frac{1}{2}}g^{rr}g_{uu,rr}-2g^{rp}g_{u[u,p],r} \nonumber\\
&&  +\textstyle{\frac{1}{2}}g^{pq}(2g_{up,uq}-g_{pq,uu}-g_{uu,pq}) \nonumber \\
&& -\textstyle{\frac{1}{2}}g^{rp}g^{rq}g_{up,r}g_{uq,r}+\textstyle{\frac{1}{2}}g^{rr}g^{pq}g_{up,r}g_{uq,r} \nonumber \\
&& +\textstyle{\frac{1}{2}}g^{pq}g^{rm}g_{up,r}(2g_{q(u,m)}-g_{um,q}) \nonumber \\
&& -\textstyle{\frac{1}{4}}g^{pq}g_{uu,r}\big[2g_{up,q}-g_{pq,u}-g^{rm}\left(2g_{mp,q}-g_{pq,m}\right)\big] \nonumber \\
&& +\textstyle{\frac{1}{2}}g^{pq}g_{up,r}\left[g_{uu,q}-g^{rm}\left(2g_{m(u,q)}-g_{uq,m}\right) \right] \nonumber \\
&& + \textstyle{\frac{1}{4}}g^{pq}g^{mn}\left(2g_{n(u,p)}-g_{up,n}\right)\left(2g_{m(u,q)}-g_{uq,m}\right)\nonumber \\
&& - \textstyle{\frac{1}{4}}g^{pq}g^{mn}\left(2g_{un,u}-g_{uu,n}\right)\left(2g_{mp,q}-g_{pq,m}\right) , \label{Ricci uu - Kundt}
\end{eqnarray}
and the Ricci scalar curvature of the general Kundt spacetime (\ref{obecny Kundtuv prostorocas}) is given by
\begin{eqnarray}
R \rovno \,^{S}\!R+g_{uu,rr}-2g^{rp}g_{up,rr}-2g^{pq}g_{up,rq} \nonumber \\
&&\ -{\textstyle\frac{3}{2}}g^{pq}g_{up,r}g_{uq,r}+g^{pq}g^{mn}g_{um,r}\,(2g_{np,q}-g_{pq,n})\, . \label{Ricci scalar - Kundt}
\end{eqnarray}
The Weyl tensor components are defined as
\begin{eqnarray}
&&
C_{abcd} = {\textstyle R_{abcd}-\frac{1}{D-2}\left(\,g_{ac}\,R_{bd}-g_{ad}\,R_{bc}+g_{bd}\,R_{ac}-g_{bc}\,R_{ad}\right)}\nonumber \\
&&\hspace{13mm}
{\textstyle+\frac{1}{(D-1)(D-2)}R\,(\,g_{ac}\,g_{bd}-g_{ad}\,g_{bc}) }\, . \label{C_abcd}
\end{eqnarray}
Notice that ${C_{rprq} = 0}$.

\section{Curvature tensors for Kundt geometries of type~II}
\label{appendixB}

In this appendix, we explicitly present the Riemann, Ricci and Weyl tensors for an arbitrary Kundt spacetime of dimension ${D\ge4}$ which is \emph{algebraically special} (of type~II or more special) \emph{with respect to} $\boldk$, the generator of a privileged null congruence with vanishing optical scalars. As in the main text, we do not apply any field equations.

From section~\ref{claasifKundtI} it follows that the most general Kundt metric of algebraic type~II with double WAND ${\boldk=\partial_r}$ can be written in the form (\ref{obecny Kundt II}) with
\begin{equation}
g_{up}=e_p(u,x)+ f_p(u,x) \,r\, , \label{guiprotypIIap}
\end{equation}
for any ${p=2,\ldots,D-1\,}$. The contravariant terms (\ref{cocontrarelations}) are thus
\begin{equation}
g^{rp} = e^p+ f^p \,r\, , \quad
g^{rr} = -g_{uu}+e^pe_p+2\,e^pf_p\, r+ f^pf_p\, r^2\, ,
 \label{guiprotypIIapcontrb}
\end{equation}
where ${e^p(u,x) \equiv g^{pq}e_q}$, ${f^p(u,x) \equiv g^{pq}f_q}$, whereas $g_{uu}(r,u,x)$ is an arbitrary function of all the coordinates. The coordinate components of the Riemann curvature tensor (\ref{Riemann rprq - Kundt})--(\ref{Riemann upuq - Kundt}) then reduce to
\begin{eqnarray}
&&\hspace{-25mm}
R_{rprq} = 0\, , \label{Riemann rprq - KundtII} \\
&&\hspace{-25mm}
R_{rpru} = 0 \, , \label{Riemann rpru - KundtII} \\
&&\hspace{-25mm}
R_{ruru} = -\textstyle{\frac{1}{2}}g_{uu,rr}+\textstyle{\frac{1}{4}}f^p f_p \, , \label{Riemann ruru - KundtII} \\
&&\hspace{-25mm}
R_{rpmq} = 0 \, , \label{Riemann rpkq - KundtII} \\
&&\hspace{-25mm}
R_{rpuq} = \textstyle{\frac{1}{2}}f_{pq}+\textstyle{\frac{1}{2}}F_{pq}\, , \label{Riemann rpuq - KundtII} \\
&&\hspace{-25mm}
R_{rupq} = F_{pq} \, , \label{Riemann rupq - KundtII} \\
&&\hspace{-25mm}
R_{ruup} = \textstyle{\frac{1}{2}}g_{uu,rp}-\textstyle{\frac{1}{2}}f_{p,u}+\textstyle{\frac{1}{4}}rf^{q}(f_{q}f_{p}-2F_{qp}) +\textstyle{\frac{1}{4}}f^q(e_qf_p-2E_{qp})\, , \label{Riemann ruup - KundtII} \\
&&\hspace{-25mm}
R_{mpnq} \,= \,^{S}\!R_{mpnq} \, , \label{Riemann kplq - KundtII} \\
&&\hspace{-25mm}
R_{upmq} = \ \,r\Big[f_{[q||m]||p}+\textstyle{\frac{1}{2}}(f_{pm}f_{q}-f_{pq}f_{m})\Big] \nonumber \\
&&\hspace{-10mm}
+e_{[q||m]||p}+\textstyle{\frac{1}{2}}(e_{pm}f_{q}-e_{pq}f_{m})+g_{p[m,u||q]}\, , \label{Riemann upkq - KundtII} \\
&&\hspace{-25mm}
R_{upuq} = -\textstyle{\frac{1}{2}}(g_{uu})_{||p||q} -\textstyle{\frac{1}{4}}g_{uu}f_pf_q-\textstyle{\frac{1}{2}}g_{uu,r}(rf_{(p||q)}+e_{pq}) +\textstyle{\frac{1}{4}}(g_{uu,p}f_q+g_{uu,q}f_p) \nonumber \\
&&\hspace{-13mm}
+r^2\Big[\textstyle{\frac{1}{4}}f_pf_qf^mf_m-\textstyle{\frac{1}{2}}f^m(F_{mp}f_q+F_{mq}f_p)+g^{mn}F_{mp}F_{nq}\Big] \nonumber \\
&&\hspace{-13mm}
+r\,\Big[f_{(p,u||q)}+\textstyle{\frac{1}{2}}f_pf_qf^{m}e_{m} -\textstyle{\frac{1}{2}}f^m(E_{mp}f_q+E_{mq}f_p)-\textstyle{\frac{1}{2}}e^m(F_{mp}f_q+F_{mq}f_p) \nonumber \\
&&\hspace{-6mm} +g^{mn}\big(E_{mp}F_{nq}+E_{mq}F_{np}\big)\Big] \label{Riemann upuq - KundtII}\\\
&&\hspace{-13mm}
+e_{(p,u||q)} -\textstyle{\frac{1}{2}}g_{pq,uu}+\textstyle{\frac{1}{4}}e^me_mf_pf_q -\textstyle{\frac{1}{2}}e^m\big(E_{mp}f_q+E_{mq}f_p\big)+g^{mn}E_{mp}E_{nq}\,, \nonumber
\end{eqnarray}
where the useful geometric quantities $f_{p||q}$, ${{f^p}_{||p}}$, $f_{pq}$, $f$, $F_{pq}$, etc. are defined in (\ref{defderiv})--(\ref{Laplaceguut}).
The symbol $||$ indicates the covariant derivative with respect to the spatial metric $g_{pq}$ in the transverse ${(D-2)}$-dimensional Riemannian space.

Similarly, from (\ref{Ricci rr - Kundt})--(\ref{Ricci uu - Kundt}) and (\ref{Ricci scalar - Kundt}), using the identity
\begin{equation}
(f^pe^q-f^qe^p)\,g_{qm,p} \equiv \Gamma^q_{pm}(e_qf^p-f_q\,e^p)\,, \label{identitef}
\end{equation}
we obtain the following coordinate components of the Ricci tensor and the Ricci scalar:
\begin{eqnarray}
R_{rr}\rovno 0\, , \label{Ricci rr - KundtII}\\
R_{rp}\rovno 0\, ,\label{Ricci rk - KundtII}\\
R_{ru}\rovno -\textstyle{\frac{1}{2}}g_{uu,rr}+{\textstyle\frac{1}{2}} f+{\textstyle\frac{1}{4}} f^p f_p\, , \label{Ricci ru - KundtII} \\
R_{pq}\rovno \!^{S}\!R_{pq}-f_{pq}\, ,
\label{Ricci pq - KundtII}\\
R_{up}\rovno -\textstyle{\frac{1}{2}}g_{uu,rp}+\textstyle{\frac{1}{2}}f_{p,u} \nonumber \\
&& +r\,\left[\, g^{mn}f_{[m||p]||n}+2f^qF_{qp}-\textstyle{\frac{1}{2}}(f+\textstyle{\frac{1}{2}}f^{q}f_{q})f_{p}\,\right] \nonumber \\
&& +\hspace{2.9mm} \big[\, g^{mn}e_{[m||p]||n}+e^qF_{qp}-\textstyle{\frac{1}{2}}g^{mn}\,e_{mn}f_{p} \nonumber \\
&& \hspace{8mm}+\textstyle{\frac{1}{2}}(f^{q}e_{q||p}-e^{q}f_{p||q}-e^qf_q f_p) +g^{mn} g_{m[p,u||n]}\,\big] , \label{Ricci uk - KundtII}\\
R_{uu}\rovno -\textstyle{\frac{1}{2}}\triangle g_{uu} +\textstyle{\frac{1}{2}} g_{uu}(\,g_{uu,rr}-f^p f_p) \nonumber\\
&& +\textstyle{\frac{1}{2}}g_{uu,p}\,f^p+(f_{p,u}-g_{uu,rp})(rf^p+e^p) \nonumber \\
&& -\textstyle{\frac{1}{2}}g_{uu,r}g^{pq}(rf_{(p||q)}+e_{pq})
-\textstyle{\frac{1}{2}}g_{uu,rr}(r^2 f^pf_p+2rf^pe_p+e^pe_p) \nonumber \\
&& +r^2\,g^{mn}g^{pq}F_{mp}F_{nq} \nonumber \\
&& +r\, \left[\,g^{pq}(f_{p,u})_{||q}+2f^p e^qF_{pq}+2\,g^{mn}g^{pq}E_{mp}F_{nq}\,\right] \nonumber \\
&& +\hspace{2.9mm} \big[\,g^{pq}(e_{p,u})_{||q}-\textstyle{\frac{1}{2}}g^{pq}g_{pq,uu}
+\textstyle{\frac{1}{2}}(f^p f_p)(e^qe_q)-\textstyle{\frac{1}{2}}(f^p e_p)^2 \nonumber \\
&& \hspace{8.3mm}+2f^p e^qe_{[p||q]}+g^{mn}g^{pq}E_{mp}E_{nq}\,\big] , \label{Ricci uu - KundtII}
\end{eqnarray}
and
\begin{eqnarray}
R \rovno g_{uu,rr}-{\textstyle\frac{1}{2}}f^p f_p+\,^{S}\!R-2f\, . \label{Ricci scalar - KundtII}
\end{eqnarray}

Finally, after straightforward but very lengthy calculation, the explicit Weyl tensor coordinate components (\ref{C_abcd}) become
\begin{eqnarray}
&&\hspace{-25mm}
C_{rprq} = 0 \, , \label{Weyl_rprqII} \\
&&\hspace{-25mm}
C_{rpru} = 0 \, , \label{Weyl_rpruII} \\
&&\hspace{-25mm}
C_{ruru} = \frac{D-3}{D-1}\left[-\frac{1}{2}g_{uu,rr}+\frac{1}{4}f^p f_p-\frac{1}{D-2}\left(\frac{\,^{S}\!R}{D-3}+f\right)\right], \label{Weyl_ruruII} \\
&&\hspace{-25mm}
C_{rpmq} = 0 \,, \label{Weyl_rpkqII} \\
&&\hspace{-25mm}
C_{rpuq} = \frac{g_{pq}}{(D-1)(D-2)}
\left[\frac{1}{2}(D-3)\,g_{uu,rr}-\frac{1}{4}(D-3)f^m f_m-\,^{S}\!R -\frac{1}{2}(D-5)f\right] \nonumber \\
&& \hspace{-1.3mm}
+\frac{1}{D-2} \left[\,^{S}\!R_{pq}+\frac{1}{2}(D-4)f_{pq} \right]+\frac{1}{2}F_{pq}\, ,\label{Weyl_rpuqII}\\
&&\hspace{-25mm}
C_{rupq} = F_{pq}\, , \label{Weyl_rupqII} \\
&&\hspace{-25mm}
C_{ruup} = \frac{D-3}{D-2}\,\bigg\{\frac{1}{2}\,g_{uu,rp}
+r\bigg[\frac{f_{p}}{D-1}\left(\frac{1}{2}\,g_{uu,rr}+\frac{1}{4}(D-2)f^mf_m-\left(\frac{\,^{S}\!R}{D-3}+f\right)\!\right) \nonumber \\
&&\hspace{21mm}+\frac{1}{D-3}\,g^{mn}f_{[m||p]||n}-\frac{1}{2}\frac{D-6}{D-3}\,f^q F_{qp}\bigg] \nonumber\\
&&\hspace{15.5mm}+\bigg[\frac{e_{p}}{D-1}\left(\frac{1}{2}\,g_{uu,rr}-\frac{1}{4}f^mf_m-\left(\frac{\,^{S}\!R}{D-3}+f\right)\!\right)  \nonumber\\
&&\hspace{21mm}-\frac{1}{2}f_{p,u}+\frac{1}{4}f^qe_qf_p-\frac{1}{2}f^qE_{qp}-\frac{1}{D-3}\,X_p\bigg]\bigg\}\, , \label{Weyl_ruupII}\\
&&\hspace{-25mm}
C_{mpnq} = \,^{S}\!R_{mpnq}-\frac{1}{D-2}\Big(g_{mn}(\,^{S}\!R_{pq}-f_{pq})-g_{mq}(\,^{S}\!R_{pn}-f_{pn}) \nonumber \\
&& \hspace{12.2mm}+g_{pq}(\,^{S}\!R_{mn}-f_{mn})-g_{pn}(\,^{S}\!R_{mq}-f_{mq})\Big)\nonumber \\
&&\hspace{-11mm}+\frac{1}{(D-1)(D-2)}\Big(g_{uu,rr}-\frac{1}{2}f^s f_s+\,^{S}\!R-2f \Big)(\,g_{mn}g_{pq}-g_{mq}g_{pn})\, , \label{Weyl_kplqII}
\\
&&\hspace{-25mm}
C_{upmq} = X_{pmq} +rf_{[q||m]||p}-F_{qm}e_p+\frac{1}{2}\big(F_{pq}e_m-F_{pm}e_q\big)\nonumber \\
&&\hspace{-11mm}
-\frac{r f_{m}+e_{m}}{D-2}\left[\left(\!\,^{S}\!R_{pq}-\frac{g_{pq}}{D-2}\,^{S}\!R\right)
+\frac{1}{2}(D-4)\left(f_{pq}-\frac{g_{pq}}{D-2}f\right)\right] \nonumber \\
&&\hspace{-11mm}+\frac{g_{pq}}{D-2}
\bigg\{\frac{1}{2}\,g_{uu,rm}+r\bigg[\frac{f_{m}}{D-1}\bigg(g_{uu,rr}+\frac{1}{4}(D-3)f^sf_s
 \nonumber \\
&&\hspace{25mm}-\frac{D-3}{D-2}\bigg(\frac{\,^{S}\!R}{D-3}+f\bigg)\bigg)-g^{ns}f_{[n||m]||s}-2\,f^n F_{nm}\bigg] \nonumber\\
&&\hspace{19mm}+\bigg[\frac{e_{m}}{D-1}\left(g_{uu,rr}-\frac{1}{2}f^sf_s-\frac{D-3}{D-2}\left(\frac{\,^{S}\!R}{D-3}+f\right)\!\right)  \nonumber\\
&&\hspace{27mm}-\frac{1}{2}f_{m,u}+\frac{1}{4}f^ne_nf_m-\frac{1}{2}f^nE_{nm} +X_m\bigg]\bigg\}\, ,
\nonumber \\
&&\hspace{-11mm}
+\frac{r f_{q}+e_{q}}{D-2}\left[\left(\!\,^{S}\!R_{pm}-\frac{g_{pm}}{D-2}\,^{S}\!R\right)
+\frac{1}{2}(D-4)\left(f_{pm}-\frac{g_{pm}}{D-2}f\right)\right] \nonumber \\
&&\hspace{-11mm}-\frac{g_{pm}}{D-2}
\bigg\{\frac{1}{2}\,g_{uu,rq}+r\bigg[\frac{f_{q}}{D-1}\bigg(g_{uu,rr}+\frac{1}{4}(D-3)f^sf_s \nonumber \\
&&\hspace{25mm}-\frac{D-3}{D-2}\bigg(\frac{\,^{S}\!R}{D-3}+f\bigg)\bigg) -g^{ns}f_{[n||q]||s}-2\,f^n F_{nq}\bigg] \nonumber\\
&&\hspace{19mm}+\bigg[\frac{e_{q}}{D-1}\left(g_{uu,rr}-\frac{1}{2}f^sf_s-\frac{D-3}{D-2}\left(\frac{\,^{S}\!R}{D-3}+f\right)\!\right)  \nonumber\\
&&\hspace{27mm}-\frac{1}{2}f_{q,u}+\frac{1}{4}f^ne_nf_q-\frac{1}{2}f^nE_{nq} +X_q\bigg]\bigg\}\, , \label{Weyl_upkqII} \\
&&\hspace{-24mm}
C_{upuq} =
\frac{1}{2}\!\left[\frac{g_{pq}}{D-2}\,\triangle g_{uu}-(g_{uu})_{||p||q}\right]\nonumber\\
&&\hspace{-11mm}
-\frac{1}{2}\!\left[\frac{g_{pq}}{D-2}\,g_{uu,m}\,f^m-\frac{1}{2}\left(g_{uu,p}\,f_q+g_{uu,q}\,f_p\right)\right] \nonumber\\
&&\hspace{-11mm}
+\frac{1}{2}g_{uu,r}\!\left[\frac{g_{pq}}{D-2}\,g^{mn}\left(r f_{(m||n)}+e_{mn}\right)-\left(r f_{(p||q)}+e_{pg}\right)\right]
\nonumber\\
&&\hspace{-11mm}
-\frac{g_{pq}}{D-2}\left(f_{m,u}-g_{uu,rm}\right)\left(r f^m+e^m\right)\nonumber\\
&&\hspace{-5mm}
+\frac{1}{D-2} \left[ \frac{1}{2}(f_{p,u}-g_{uu,rp})(r f_q+e_q)+\frac{1}{2}(f_{q,u}-g_{uu,rq})(r f_p+e_p)\right]\nonumber\\
&&\hspace{-11mm}
+\frac{g_{pq}\,g_{uu}}{(D-1)(D-2)}\left[-\frac{1}{2}(D-3) g_{uu,rr}+\frac{1}{2}(D-2)f^mf_m+\,^{S}\!R-2f\right]\nonumber\\
&&\hspace{-5mm}
-\frac{g_{uu}}{D-2}\left[\frac{1}{4}(D-2)f_pf_q+\,^{S}\!R_{pq}-f_{pq}\right]
\nonumber\\
&&\hspace{-11mm}
+\frac{g_{pq}}{D-2}\,g_{uu,rr}\,\frac{1}{2}\left(r^2 f^m f_m+2\,rf^m e_m+e^me_m\right)\nonumber\\
&&\hspace{-11mm}
-\frac{1}{(D-1)(D-2)}\left[g_{uu,rr}-\frac{1}{2}f^mf_m+\,^{S}\!R-2f\right]
\nonumber\\
&&\hspace{42mm}
\times\Big(r^2 f_p f_q+r(f_p e_q+f_qe_p)+e_pe_q\Big)
\nonumber\\
&&\hspace{-11mm}
+ r^2 \Bigg[\, g^{mn}F_{mp}F_{nq} -\frac{g_{pq}}{D-2}\,g^{mn}g^{st}F_{ms}F_{nt}\nonumber\\
&&\hspace{-5mm}
-\frac{1}{2}\frac{D-6}{D-2}\,f^m(F_{mp}\,f_q+F_{mq}\,f_p)
+\frac{g^{mn}}{D-2}\,(f_{[m||p]||n}\,f_q+f_{[m||q]||n}\,f_p)\nonumber\\
&&\hspace{-5mm}
+\frac{1}{D-2}\left(\frac{1}{4}(D-4)f^mf_m-f\right)\!f_pf_q\Bigg]   \nonumber\\
&&\hspace{-11mm}
+ r\,\,\Bigg[\, f_{(p,u||q)}-\frac{g_{pq}}{D-2}\,g^{mn} f_{(m,u||n)} -\frac{1}{D-2}\left(X_p f_q+X_q f_p\right) \nonumber\\
&&\hspace{-5mm}
+g^{mn}\left(E_{mp}F_{nq}+E_{mq}F_{np}\right)-2\frac{g_{pq}}{D-2}\,g^{mn}g^{st}E_{ms}F_{nt} \nonumber\\
&&\hspace{-5mm}
-\frac{1}{2}\frac{D-3}{D-2}\,f^m(E_{mp}f_q+E_{mq}f_p)+\frac{g^{mn}}{D-2}\,(f_{[m||p]||n}\,e_q+f_{[m||q]||n}\,e_p)\nonumber\\
&&\hspace{-5mm}
+ \frac{2}{D-2}\,f^m\,[(F_{mp}\,e_q+F_{mq}\,e_p)-g_{pq}F_{mn}\,e^n]\nonumber\\
&&\hspace{-5mm}
- \frac{1}{2}\,e^m(F_{mp}\,f_q+F_{mq}\,f_p)\nonumber\\
&&\hspace{-5mm}
+\frac{1}{2}\frac{D-3}{D-2}\,f^me_mf_pf_q
-\frac{1}{D-2}\left(\!f+\frac{1}{4}f^mf_m\right)\!\left(f_pe_q+f_qe_p\right)\Bigg]  \nonumber\\
&&\hspace{-11mm}
+ \ \Bigg[ \big(e_{(p,u||q)}-{\textstyle\frac{1}{2}}g_{pq,uu}\big)-\frac{g_{pq}}{D-2}\,g^{mn}\big(e_{(m,u||n)}-{\textstyle\frac{1}{2}}g_{mn,uu}\big)\nonumber\\
&&\hspace{-5mm}
-\frac{1}{D-2}\left(X_p e_q+X_q e_p\right)+g^{mn}E_{mp}E_{nq}-\frac{g_{pq}}{D-2}\,g^{mn}g^{st}E_{ms}E_{nt} \nonumber\\
&&\hspace{-5mm}
- \frac{1}{2}\,e^m(E_{mp}f_q+E_{mq}f_p)+\frac{1}{2}\frac{1}{D-2}\,f^m(E_{mp}\,e_q+E_{mq}\,e_p)\nonumber\\
&&\hspace{-5mm}
-2\frac{g_{pq}}{D-2}\,f^me^n \,e_{[m||n]}+\frac{1}{2}\frac{1}{D-2}\,f^me_m\left( g_{pq}f^ne_n-\frac{1}{2}(f_pe_q+f_qe_p)\right) \nonumber\\
&&\hspace{-5mm}
-\frac{1}{2}\,e^me_m\left(\frac{g_{pq}}{D-2}f^nf_n-\frac{1}{2}f_pf_q\right)-\frac{1}{D-2}\,f\, e_pe_q\Bigg]
\, , \label{Weyl_upuqII}
\end{eqnarray}
in which $X_{pmq}$ is given by (\ref{Xipj*}), and $X_q$ by (\ref{Xp*}).

\end{document}